%% file: main.tex
\documentclass[sigconf]{acmart}
\usepackage{booktabs} % For formal tables
\usepackage{graphicx,multicol,latexsym,amsmath,amssymb}
\usepackage[labelformat=simple]{subfig}
\usepackage[toc,page]{appendix}
\acmDOI{unknown}

\acmISBN{123-4567-24-567/08/06}

\acmConference[MLG'2018]{Workshop on Mining and Learning with Graphs}{August 2018}{London, UK} 
\acmYear{2018}
\copyrightyear{2018}

\begin{document}
%\title{Tracking and modelling of disease spreading\\ using dynamic individual contacts from social networks}
\title{A Graph Model with Indirect Co-location Links}

% \titlenote{Produces the permission block, and
%   copyright information}w
% \subtitle{Extended Abstract}
% \subtitlenote{The full version of the author's guide is available as
%   \texttt{acmart.pdf} document}

\author{Md Shahzamal}
% \authornote{Dr.~Trovato insisted his name be first.}
% \orcid{1234-5678-9012}
\affiliation{
  \institution{Macquarie University}
%   \streetaddress{P.O. Box 1212}
  \city{Sydney} 
  \state{Australia} 
%   \postcode{43017-6221}
}
\email{md.shahzamal@students.mq.edu.au}

\author{Raja Jurdak}
\affiliation{%
  \institution{CSIRO}
  \city{Brisbane} 
  \state{Australia} 
%   \postcode{43017-6221}
}
\email{raja.jurdak@data61.csiro.au}

\author{Bernard Mans}
% \authornote{This author is the
%   one who did all the really hard work.}
\affiliation{%
  \institution{Macquarie University}
%   \streetaddress{Sydney}
  \city{Sydney} 
  \country{Australia}}
\email{bernard.mans@mq.edu.au}

\author{Frank de Hoog}
\affiliation{%
  \institution{CSIRO}
  \city{Canberra}
  \country{Australia}
}
\email{frank.dehoog@data61.csiro.au}
% \author{Aparna Patel} 
% \affiliation{%
%  \institution{Rajiv Gandhi University}
%  \streetaddress{Rono-Hills}
%  \city{Doimukh} 
%  \state{Arunachal Pradesh}
%  \country{India}}
% \author{Huifen Chan}
% \affiliation{%
%   \institution{Tsinghua University}
%   \streetaddress{30 Shuangqing Rd}
%   \city{Haidian Qu} 
%   \state{Beijing Shi}
%   \country{China}
% }

% \author{Charles Palmer}
% \affiliation{%
%   \institution{Palmer Research Laboratories}
%   \streetaddress{8600 Datapoint Drive}
%   \city{San Antonio}
%   \state{Texas} 
%   \postcode{78229}}
% \email{cpalmer@prl.com}

% \author{John Smith}
% \affiliation{\institution{The Th{\o}rv{\"a}ld Group}}
% \email{jsmith@affiliation.org}

% \author{Julius P.~Kumquat}
% \affiliation{\institution{The Kumquat Consortium}}
% \email{jpkumquat@consortium.net}

% % The default list of authors is too long for headers.
% \renewcommand{\shortauthors}{B. Trovato et al.}

\begin{abstract}
Graph models are widely used to analyse diffusion processes embedded in social contacts and to develop applications. A range of graph models are available to replicate the underlying social structures and dynamics realistically. However, most of the current graph models can only consider concurrent interactions among individuals in the co-located interaction networks. They do not account for indirect interactions that can transmit spreading items to individuals who visit the same locations at different times but within a certain time limit. The diffusion phenomena occurring through direct and indirect interactions is called same place different time (SPDT) diffusion. This paper introduces a model to synthesize co-located interaction graphs capturing both direct interactions, where individuals meet at a location, and indirect interactions, where individuals visit the same location at different times within a set timeframe. We analyze 60 million location updates made by 2 million users from a social networking application to characterize the graph properties, including the space-time correlations and its time evolving characteristics, such as bursty or ongoing behaviors. The generated synthetic graph reproduces diffusion dynamics of a realistic contact graph, and reduces the prediction error by up to $82\%$ when compared to other contact graph models, thus demonstrating its potential for forecasting epidemic spread.
\end{abstract}

% \begin{abstract}
% This paper provides a sample of a \LaTeX\ document which conforms,
% somewhat loosely, to the formatting guidelines for
% ACM SIG Proceedings.\footnote{This is an abstract footnote}
% \end{abstract}

%
% The code below should be generated by the tool at
% http://dl.acm.org/ccs.cfm
% Please copy and paste the code instead of the example below. 
%

% \ccsdesc[300]{Computing methodologies~Modeling and simulation}


\maketitle
\input{introduction}
\input{network}
\input{modelfit}
\input{netproperties}
\input{diffussion}
\input{discussion}
\input{apendix}
\bibliographystyle{IEEEtran}
\bibliography{sigproc} 

\end{document}

%% file: introduction.tex
\section{Introduction}
Modelling diffusion processes driven by social contacts have recently received significant research attention. These processes range from viral marketing in on-line social networks to infectious disease spreading in social contact within populations. Studying these processes in large real scenarios is not possible without detailed information about the contact patterns and their timing behaviors. However, gathering such large scale data is expensive and complex due to collection methods and privacy concerns. The alternative method is to develop synthetic network structures capturing the statistical properties of real contacts. This synthetic structure is represented by graphs where individuals are presented as actors (i.e., nodes) and relationships among them as edges (i.e., links). In the traditional approach, the edges between nodes are static. However, there are several types of contacts that are not permanent over time such as physical contacts between individuals. 

To capture the temporal dynamics of interactions, various dynamic contact graphs have been introduced in the literature~\cite{holme2015modern,kim2011modeling,vazquez2003growing,perra2012activity}.
% In our study, we focus on dynamic graphs created by co-located interactions among individuals and spread infectious items to nodes present in the graph.
% Developing graph models follows two mechanisms. Some models are mechanistic applying simple rules to generate connectivity in the graph. These graphs are easily mathematically tractable. Another stream uses the statistical approaches to generate graph structure.
Dynamic contact graphs are frequently generated using statistical methods. These models provide realistic graphs for hypothesis testing, "what-if" scenarios, and simulations, but are often mathematically untraceable. Dynamic models require to make graph entities and edges evolve over time and simultaneously maintain the underlying social structures~\cite{kim2011modeling}. The preferential attachment is used to study dynamic graphs: new nodes join the graph with non-uniform probability~\cite{vazquez2003growing}. But, it cannot help maintain social structures. This is addressed by the Nearest Neighbor Model %~\cite{} 
which connects two nodes if they have common neighbor nodes. Another model called Homophily~\cite{mcpherson2001birds} connects two nodes having common interests to form community structures. These models are connectivity driven and face difficulties to capture features of real graphs of large size. The authors of~\cite{perra2012activity} proposed activity driven dynamic network models to build interaction graphs where a node activates at a given time, with its potentiality, and starts creating links with other nodes. This basic activity driven network model (BADN) has been upgraded to capture realistic graph properties applying preferential attachments, reinforcement procedures and attractiveness~\cite{karsai2014time}.
Current co-located interaction models, however, assume that links between two individuals are created when they are both present at the same location. Thus, the infectious items are transmitted through a link when both infected and susceptible individuals are present~\cite{holme2015modern,baccelli2013multi}. We refer to diffusion due to these individual to individual level transmissions as same place same time transmission (SPST) based diffusion and created contact graphs as SPST interactions graphs.
% For instance, current disease diffusion models require two individuals to be simultaneously located in the same place for a transmission to occur. 
This focus on concurrent presence, is not sufficiently representative of a class of diffusion scenarios where transmissions can occur with indirect interactions, i.e. when there is a time gap between the departure of one individual and the arrival of another. Airborne disease transmission is one such example. An infected individual can release infectious particles in the air through coughing or sneezing. These particles are then suspended in the air and an individual arriving after the departure of the infecter can still get infected~\cite{fernstrom2013aerobiology,shahzamal2017airborne}. 
%To capture such diffusion process, a diffusion model called same place different time transmission (SPDT) based diffusion has been introduced in our previous work.
%of ~\cite{}. 
In this scenario, current graph models that exclusively track concurrent interactions (SPST) can miss significant spreading events during indirect interactions, thus underestimating the diffusion dynamics. There is a need for a novel dynamic contact graph model to study SPDT diffusions.

This paper proposes a temporal graph model that considers both concurrent (direct) interactions and delayed (indirect) interactions among individuals in forming the links. We termed this model as same place different time interactions (SPDT) graph. In this graph, possible disease transmission links are created if a host node visits a location where at least one other node is present. The node's stay duration at a location with potential to spread is termed as an active period. In order to represent both direct and indirect transmission opportunities, we define the concept of an active copy of a node which is created for each active period of a node. In the proposed model, links are created between the active copies and neighbor nodes. The active copy survives for the active period, when the host is present, in addition to an indirect transmission period, when the host leaves yet the spreading items persist at the location. Thus, the SPDT graph evolves according to temporal changes of links and node's status. SPDT graph generation methods are developed using statistical distributions which are fitted with real graphs of 2 millions users. The model is then validated through its ability to reproduce diffusion dynamics of real contact graphs and studying performances compared to SPST and BADN graph models

%The remainder of this paper is organized as follows. 
The SPDT graph model is introduced in Section 2 while the graph generation is explained in Section 3. Model fitting techniques are presented in Section 4. Section 5 describes the validation of proposed model while Section 6 discusses the results and concludes the paper. 

%% file: network.tex
\section{SPDT Graph Model}
%We propose a new graph model capturing indirect transmission links occurring in co-located interactions among individuals. 
In this section, we first analyse the modeling scenarios and then describe the proposed graph model. 
\subsection{Modeling Scenarios}
\begin{figure}
    \vspace{-1.0 em}
	\centering
\includegraphics[width=0.90\linewidth, height=5.1 cm]{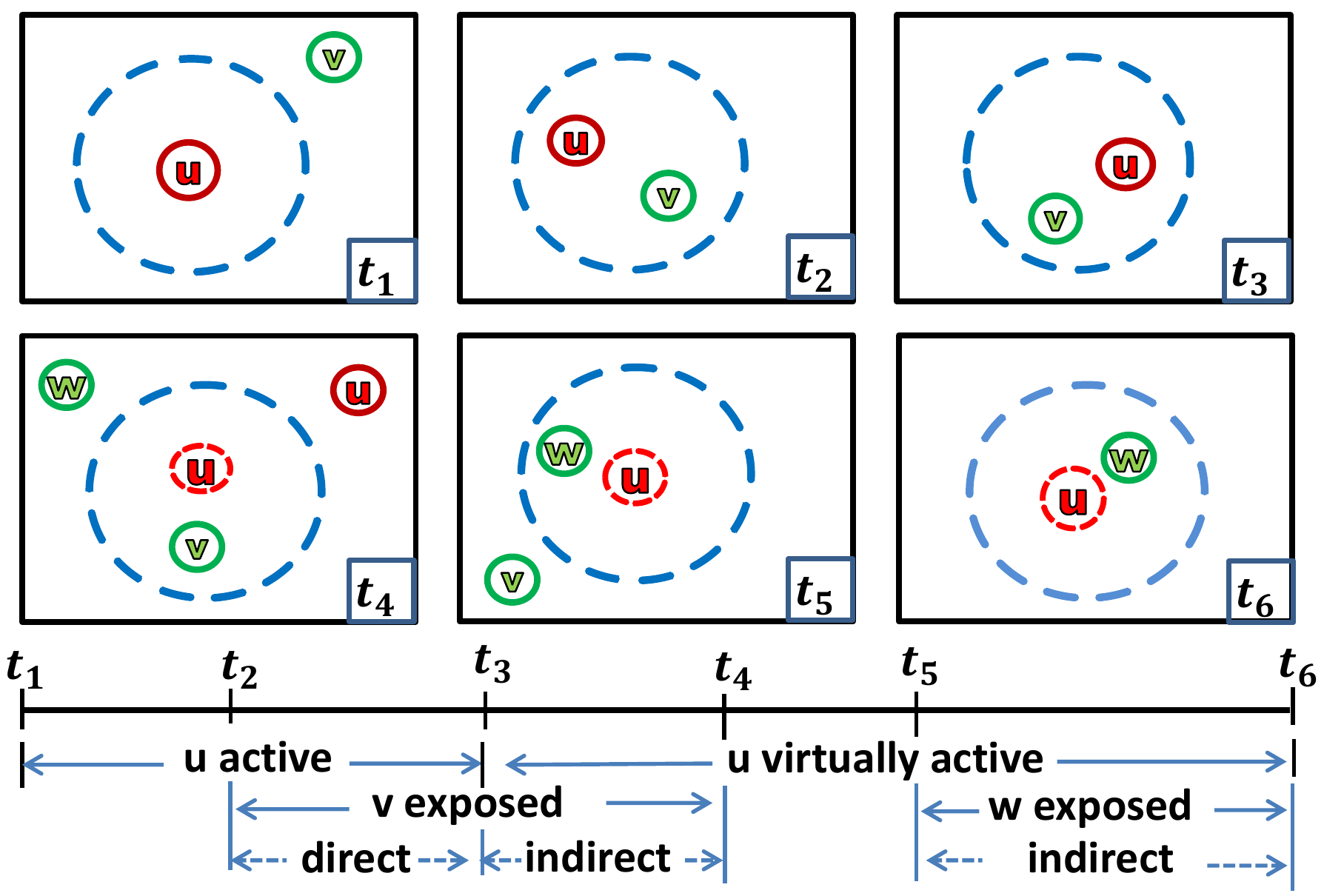}
    \vspace{-1.0 em}
	\caption{Modeling scenery}
	\label{fig:spdtp}
    \vspace{-1.5 em}
\end{figure}

We first explain the link creation process in SPDT graphs through airborne disease spreading where infected individuals deposit infectious particles at locations they have visited. These particles persist in the environment and can be transferred to susceptible individuals who are currently present nearby (direct transmission) as well as to individuals who visit the location later on (indirect transmission)~\cite{fernstrom2013aerobiology,shahzamal2017airborne}. Figure~\ref{fig:spdtp} illustrates a series of snapshots over time when different individuals visit a location $L$ (dashed circle) to create these links. Here, an infected individual $u$ (host individual) arrives at $L$ at time $t_{1}$ followed by a susceptible individual $v$ at time $t_{2}$. The appearance of $v$ at $L$ creates a directed link for transmitting disease from $u$ to $v$ and lasts until time $t_{4}$ including direct contact during $[t_{2},t_{3}]$ and indirect contact during $[t_{3},t_{4}]$. The indirect contact is created as the impact of $u$ persists (as shown by the dashed circle surrounding $u$), due to the survival of the airborne infected particles in the air. Another susceptible user $w$ arrives at $L$ at time $t_{5}$ and a link is created from $u$ to $w$ through the indirect route due to $u$'s infectious particles still being active at $L$. These links are called SPDT links and created through space and time. SPDT links may have components: direct and/or indirect transmission links. However, visits of infected individuals to the locations with no susceptible individuals do not lead to transmission of disease. 

\subsection{Model Definition}
Individuals may move to various places during their infectious periods while traveling to public places such as office, school, shopping malls and bus stations and visiting tourist places. However, only visits or stays of infected individuals to locations where other susceptible individuals are present create disease transmission links. Our goal is to develop a graph model that is capable of capturing the indirect transmission links along with direct links for co-located interactions among individuals. We focus our modeling of link creation on the time domain to ensure scalability, while abstracting spatial aspects implicitly. Temporal modeling is sufficient to identify the nodes participating in possible disease transmission links. Thus, link creation events in the proposed scenario can be represented as a process where an infected individual activates for a period of time (staying at a location with susceptible individuals) and creates SPDT links. Then, the infected individual becomes inactive for a period of time during which he does not create SPDT links. Inactive periods represent the waiting time between two active periods. Thus, the co-located interaction status of an infected individual can be given by a set $\{a_1,w_1,a_2,w_2,.\}$ where $a$ is active and $w$ is inactive period.

We define SPDT graph as $G=(V, A, E, T)$ to represent all possible disease transmission links among nodes, where $V$ is the set of nodes. The number of nodes in the graph is constant; however, nodes may have one or more active copies in the graph which captures their ability to spread diseases both at locations they are present and at locations from which they recently departed. The set of active copies for all nodes is represented by $A$. $E$ is the set of links in the graph. The graph is represented over a discrete time set T=$\{t_1,t_2,\ldots t_z \}$. Each node in the graph creates a set of active and inactive periods $\{a_1,w_1,a_2,w_2 \ldots\}$. We define an active copy $v_{i}=v(t_s^{i},t_l^{i})$ for an active period $a_i$ of node $v$, where $a_i$ starts at time step $t_s^{i}$ and finishes at $t_l^{i}$. Thus, each node will have several such temporal copies for the observation period. For an active copy $v_{i+1}$ of a node $v$, $t_{s}^{i+1}$ should be greater than $t_{l}^{i}$ of $v_i$ to capture the requirement that a node should have left the first location before arriving in another location. In this graph, a link $e_{vu}\in E$ is defined between an active copy $v_i$ of host node $v$ and neighbor node $u$ (node $u$ visits the current or recent location of node $v$) as $e_{vu}=(v_i,u,t_{s}^{\prime},t_{l}^{\prime})$ where $t_s^{\prime}$ is the joining time and $t_l^{\prime}$ is departure time of $u$ from the interacted location. The value of $t_s^{\prime}$ should be within $t_s^{i}$ and $t_l^{i}+\delta$ where $\delta$ is the time period allowed to create indirect transmission links. Thus, an active copy $v_i$ of a node v expires after $t_l^{i}+\delta$, where $\delta$ captures the decaying probability of infection after $v$ departs. During the indirect transmission period $\delta$ node v can start another active period at another location (see Fig.\ref{fig:virtual}). However, if the infected node $v$ leaves and returns to the location of $u$ within a time period $\delta$, then there will be two active copies of $v$, each with a link to the susceptible node $u$. The first copy is due to the persistent of particles from $v$'s last visit, while the second copy is due to $v$'s current visit. 

\subsection{Graph Evolution}
\begin{figure}
\vspace{-1.4 em}
	\subfloat[Active and inactive periods with underlying states]{\label{fig:states}\includegraphics[width=0.99\linewidth, height=1.3cm]{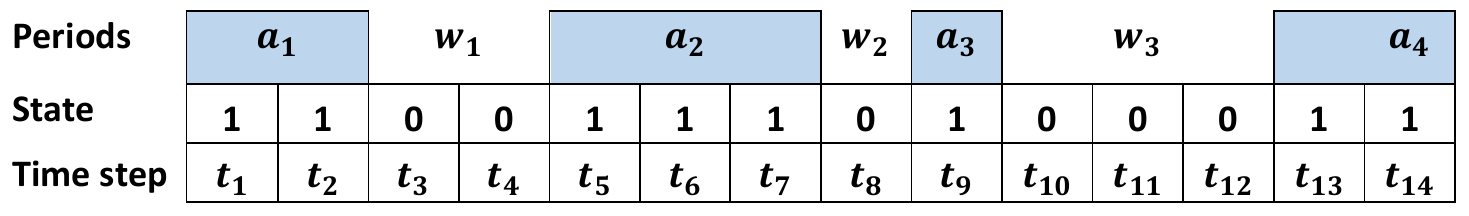}}\\
    \vspace{-0.8 em}
	 \subfloat[Active periods with the corresponding indirect periods]{\label{fig:virtual}\includegraphics[width=0.99\linewidth, height=1.50cm]{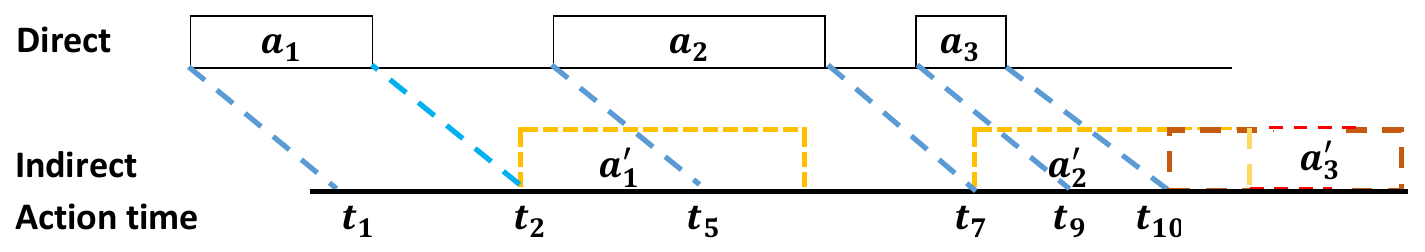}}
     \vspace{-0.6 em}
    \caption{Proposed model definition}
	\label{fig:timing}
	\vspace{-1.9 em}
\end{figure}

The evolution of the proposed graph is governed by two dynamic processes: 1) switching of nodes between active and inactive states, as in Figure~\ref{fig:states}; and 2) link creation and deletion for active copies of nodes. The total number of time steps a node remains in one state determines the current active or inactive period, leading to a set of alternating active and inactive periods $\{a_1,w_1,a_2,w_2 \ldots\}$ for an observation period (see Fig.\ref{fig:states}). As stay times at locations are not fixed~\cite{cheng2018grouping}, we define a transition probability $\rho$ to determine switching from active state to inactive state (modeling stay and departure events of a node at a location). This induces variable lengths of active periods. Similarly, the transition probability $q$ determines when a node switches from inactive to active state (modeling arrival of a node at location). A similar approach is taken to define link update dynamics in the graph. An active copy of a node creates a link to a newly arriving neighbor node with probability $p_{c}$ at each time step until it expires. We define an activation degree probability characterized by $P(d)$ to model the arrival of multiple new neighbors for an active copy. The created links break (neighbor node leaves interaction area) with probability $p_{b}$ at each time step. 

\section{Graph Generation}
We turn our attention now to develop methods for generating the proposed graph, which are designed to capture the statistical properties of realistic scenarios, social contact dynamics as well as temporal dynamics of SPDT interactions. The graph evolves with generating active copies of each node and creating their links.

% \subsection{Node Initialization}
% First, we define the initial states of nodes. In our model, states of a node are switched based on the following transition matrix: %(matrix is not working in latex and will fix later)
% \begin{equation*}
% P=\begin{bmatrix}
% q & 1-q\\
% \rho & 1-\rho
% \end{bmatrix}
% \end{equation*}
% % \[P=\mid p_i \qquad (1-p_i) \mid \]
% % \[\quad \mid \rho \qquad (1-\rho) \mid  \]
% The model follows a two state Markov-process. Thus, we need to make sure that the probability of staying at each state for a node reaches the equilibrium condition. This is obtained by assigning the probability of assuming a state by a node at the first time step as:
% \begin{equation*}
% \pi=\pi P
% \end{equation*}
% where $\pi=[\pi_{0},\pi_{1}]$, $\pi_{0}$ is the probability that the node is in inactive state $0$ and $\pi_{1}$ is the probability that node assumes active state $1$. We can represent $\pi_{0}$ and $\pi_{1}$ in terms of $\rho$ and $q$ by the following equations:
% \[\pi_{0}=(1-q)\pi_{0}+\rho \pi_{1} %\]
% % \[
% \mbox{\ \ \ \ \  and \ \ \ \ \ } 
% \pi_{1}=(1-\rho)\pi_{1}+q\pi_{0}\]
% Therefore, we obtain:
% \begin{equation}\label{son}
% \pi_{0}=\frac{\rho }{q+\rho}
% %\end{equation}
% \mbox{\ \ \ \ \  and \ \ \ \ \ }
% %\begin{equation}\label{soff}
% \pi_{1}=\frac{q}{q+\rho}
% \end{equation}
% The values of $\pi_{0}$ and $\pi_{1}$ are constant at each time step and can determine the initial state of a node. 

\subsection{\textbf{Node Activation}}
Active copies of nodes are created over time according to the active periods which are the building blocks of the graph along with nodes. Thus, we need to generate active periods and intervening inactive periods. In our model, determining whether a node will stay in the current state or transit into the other state at the next time step resembles a Bernoulli process of two outcomes. Thus, the number of time steps a node stays in a state can be obtained from a geometric distribution. With the transitional probability $\rho$ of switching from active to inactive state, the active period durations $t_a$ can be drawn from the following distribution as:
\begin{equation}\label{aprds}
Pr(t_{a}=t)=\rho (1-\rho)^{t-1}
\end{equation}
where $t=\{1,2, \ldots \}$ are the number of time steps. Similarly, the inactive period durations, $t_w$, with the transition probability $q$ can be drawn from the following distribution as:
\begin{equation}\label{iaprds}
Pr(t_{w}=t)=q (1-q)^{t-1}
\end{equation}
where $t=\{1,2, \ldots \}$ are the number of time steps. 

Now, we need to define the initial states of nodes to process active copy generation. Our model follows a two state Markov-process with transition matrix
\begin{equation*}
P=\begin{bmatrix}
q & 1-q\\
\rho & 1-\rho
\end{bmatrix}
\end{equation*}
for which the equilibrium probabilities that the node is in inactive state and active state are $\pi_0$ and $\pi_1$ respectively, where
\begin{equation}\label{son}
\pi_{0}=\frac{\rho }{q+\rho}
%\end{equation}
\mbox{\ \ \ \ \  and \ \ \ \ \ }
%\begin{equation}\label{soff}
\pi_{1}=\frac{q}{q+\rho}
\end{equation}
If the initial state of node $v$ is active, the first active copy $v_1$ is created for the time interval $(t_s^{1}=0,t_l^{1}=t_a)$. Otherwise, $v_1$ will be created for the interval $(t_s^{1}=t_w,t_l^{1}=t_w+t_a)$. Active copy creation continues over the observation period and the corresponding interval $(t_s,t_l)$ is defined according to the drawn $t_a$ and $t_w$. Active copies are generated for each node independently. The values of $\rho$ and $q$ are the same for all nodes which are fitted with real data.

\subsection{Activation Degree}
Now, we need to define interactions of neighbor nodes with an active copy. Multiple neighbor nodes can contact with an active copy. We note the number of neighbor nodes interacting with an active copy as activation degree $d$. The value of $d$ depends on the spatio-temporal dynamics of the graph and are drawn from a geometric distribution (Eq.~\ref{actdgr}) instead of finding the arrival times of neighbor nodes.  
\begin{equation} \label{actdgr}
Pr\left (d=k\right )=\left(1-\lambda\right)\lambda^{k-1}
\end{equation}
where $k=\{1,2,\ldots\}$ and scaling parameter $\lambda$. However, individuals in reality have heterogeneous accessibility to public places~\cite{alessandretti2017random} and hence activation degrees vary for individuals. Thus, heterogeneous $\lambda$ are selected for nodes and are drawn from a power law distribution of Equation~\ref{accesb}:
\begin{equation} \label{accesb}
f\left( \lambda_{i}=x\right)= \frac{\alpha x^{-(\alpha+1)}}{\xi^{-\alpha}-\psi^{-\alpha}}
\end{equation}
where $\alpha$ is the scaling parameter, $\xi$ is the lower limit of $\lambda$ and $\psi$ is the upper limit which is approximately 1. The value of $\lambda_{i}$ defines the range of variations of $d$ for active copies of a node $i$ and Equation~\ref{actdgr} ensures wide ranges for large values of $\lambda$. Combining geometry and power law distributions can generate more realistic degree distribution~\cite{chattopadhyay2014fitting} which are shown in the model fitting section.
% Then, we assign link creation delays $t_c$, time gap between arrivals of host node and neighbor node $(t_s-t_{s}^{\prime})$, and link duration $t_d$, the stay time of neighbor at the interacted location, to each link.

\subsection{Link Creation}
With the activation neighbor set, we need to define the arrival and departure dynamics of neighbor nodes for each link created with an active copy. We adopt a similar approach to the definition of active and inactive periods created with transition probabilities. We assume that each link is created with probability $p_{c}$ at each time step during the life period $(t_s,t_{l}+\delta)$ of an active copy and is broken with probability $p_{b}$ after creation. For the link creation delay $t_c$, time gap between arrivals of host node and neighbor node $(t_s-t_{s}^{\prime})$, we use the truncated geometric distribution: 
\begin{equation} \label{ldelay}
P\left (t_{c}=t\right )=\frac{p_{c} \left(1-p_{c}\right)^{t}}{1 -(1-p_{c})^{t_{a}+\delta}} 
\end{equation}
where $t=\{0, 1,2, \ldots, t_a+\delta \}$ are the number of time steps and $t_a$ is the active period duration of corresponding active copy. Truncation ensures that links are created within $t_{l}+\delta$, i.e. before the active copy expires. In contrast, the link duration $t_{d}$, the stay time of neighbor at the interacted location, does not have a specific upper bound and is generated for each link upon creation through a geometric distribution: 
\begin{equation} \label{ldur}
P\left(t_{d}=t\right)=p_{b} \left(1-p_{b} \right)^{t-1}
%P\left ( t_{d}=t \right )=\rho \left ( 1-\roh \right )^{t-1}
\end{equation}
where $t=\{1,2, \ldots \}$ are the number of time steps. For simplicity, we set $p_b=\rho$ as both probabilities relate to how long nodes stay at a location. For each link with an active copy, thus, we can find $t_{s}^{\prime}=t_s+t_c$ and $t_{l}^{\prime}=t_s^{\prime}+t_d$. A link with $t_s^{\prime}\geq t_l$ is an indirect transmission only component. Link can also have indirect component if $t_s^{\prime}< t_l > t_l^{\prime}$. The above graph generation steps capture the temporal behavior of SPDT links. The social mixing patterns are integrated by selecting the neighboring node, as we describe below.

\subsection{\textbf{Social Structure}}
Social network analysis has shown that the neighbor selection for creating a link follows a memory-based process. Thus, we apply the reinforcement process~\cite{karsai2014time} to realistically capture the repeated interactions between individuals. In this process, a neighbor node from the set of already contacted nodes is selected with probability $P(n_t+1)=n_t/(n_t+\eta)$ where $n_t$ is the number of nodes the host node already contacted up to this time $t$ and $\eta$ is the tendency to broaden the contact set size. On the other hand, a new neighbor node is selected with the probability $1-P(n_t+1)$. Besides, when a node $j$ is chosen as a new neighbor by node $i$, it is selected with the probability proportional to its $\lambda_j$ as nodes with higher $\lambda$ will be neighbors to the more nodes~\cite{alessandretti2017random}. This ensures nodes with higher potential to create links also have higher potential to receive links. 

%% file: modelfit.tex
\section{Model Fitting}
We now focus on tuning the model parameters to make them representative of real contact dynamics. While high quality empirical movement and contact data are difficult to obtain, recent location-based applications create opportunities to gather individual-level geo-tagged updates to serve as a proxy for individual movements. Here, we use the location updates from a social networking application called Momo to estimate model parameters and validate. 
% In the remainder of this section, we extract the SPDT features from the network built by Momo users from one city (Shanghai) and estimate the model parameters through our developed model fitting methods. In the following section, we validate the network properties of the estimated model parameters on another city (Bejing). 

\subsection{Data Set}
We analyze 60 million location updates collected over 32 days from 2 million Momo users of two cities (Beijing and Shanghai). The collection system retrieved location updates from the server every 15 minutes. Each update includes spatial coordinates and update times~\cite{chen2013and}. We build SPDT graphs using these updates where SPDT links are formed assuming airborne disease transmission mechanisms for co-located interactions among users. 

Consecutive updates, $\{(x_{1},t_{1}),(x_{2},t_{2}),\ldots\}$ where $x_{i}$ are the co-ordinate values and $t_{i}$ are the update times, from a user $v$ within a radius of 20m (travel distance of airborne infection particles~\cite{fernstrom2013aerobiology,shahzamal2017airborne}) of the initial update's location $x_{1}$ are indicative of the user staying within the same proximity of $x_1$. We set the threshold for time difference of any two updates to 30 minutes to remain within the same proximity, as longer gaps may indicate a data gap in the user pattern. For user $u$, its visit to the proximity of $x_{1}$ will represent an active period that creates an active copy of $v$ if a susceptible user $u$ has location updates starting at $t^{'}_{1}$ while $v$ is present, or within $\delta$ seconds after $v$ leaves the area. The user $u$ should have at least two updates within 20m of $x_{1}$ to be valid to ensure that it is in fact staying at the same proximity, and therefore can be exposed to the infectious particles, rather than simply passing by. An active period is made with duration $t_{a}=t_{k}-t_{1}$ which creates an active copy $v(t_{1},t_{k})$ of user $v$, where $t_1$ is the starting of active period and $t_k$ is the end time. If $u$'s last update within 20m around $x_{1}$ is $(x^{'}_{j},t^{'}_{j})$, an SPDT link $e_{vu}=\left(v(t_{1},t_{k}),u,t_{1}^{\prime},t_{j}^{\prime}\right)$ is created with a link creation delay $t_{c}=t^{'}_{1}-t_{1}$ and link duration $t_{d}=t^{'}_{j}-t^{'}_{1}$ for $v(t_{1},t_{k})$. Setting indirect transmission period $\delta$ to 3 hours (maximum time infectious particles can persist at a location after an infected individual leaves~\cite{fernstrom2013aerobiology}) and processing 60M location updates, we can extract about 3.4M SPDT links that are used for parameter estimation and model validation in the remainder of this section.

\subsection{Parameter Estimation}
\begin{figure}
\vspace{-1.5 em}
	\subfloat{\includegraphics[width=0.49\linewidth, height=2.6cm]{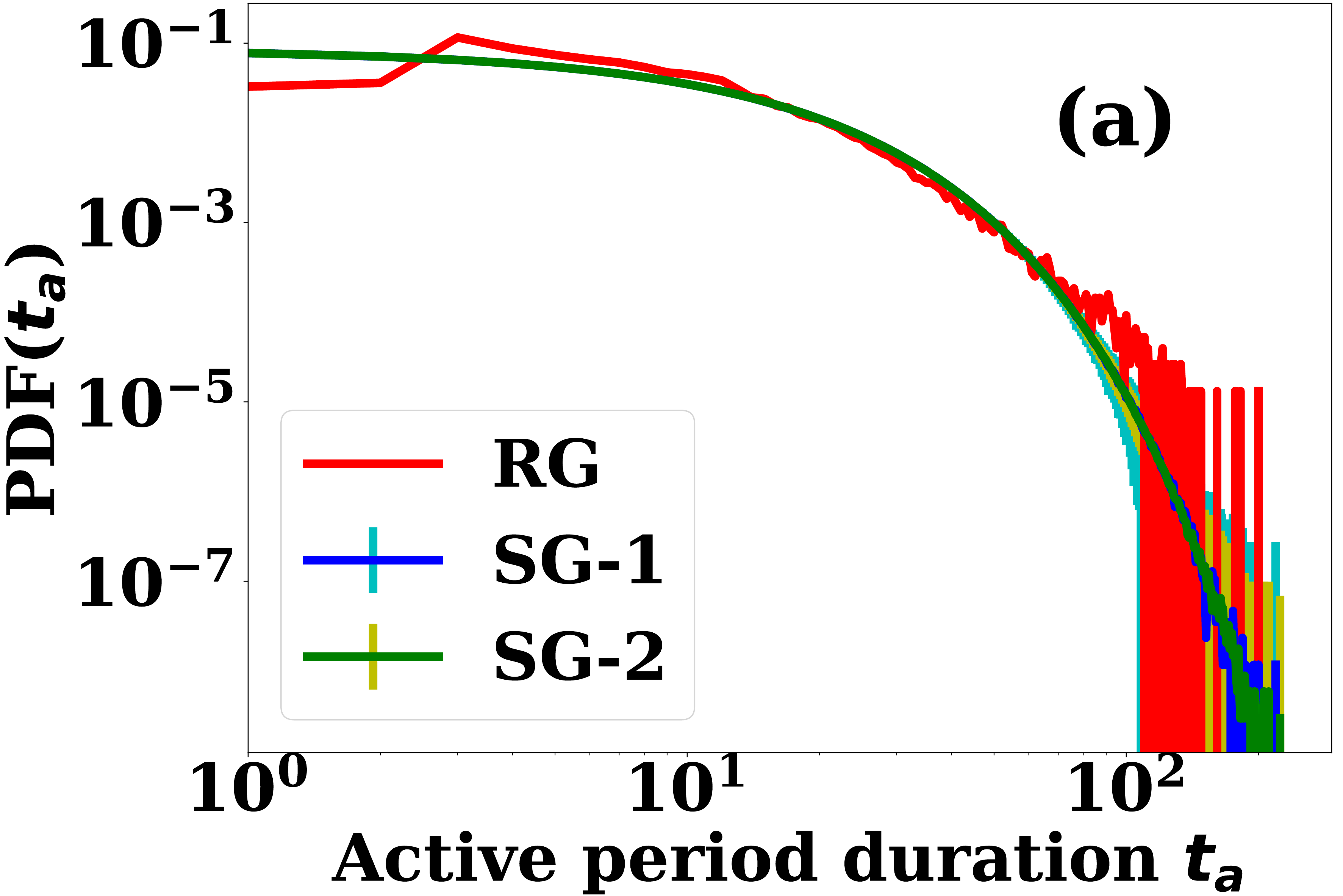}\label{fig:fitta}}~
	 \subfloat{\includegraphics[width=0.49\linewidth, height=2.6cm]{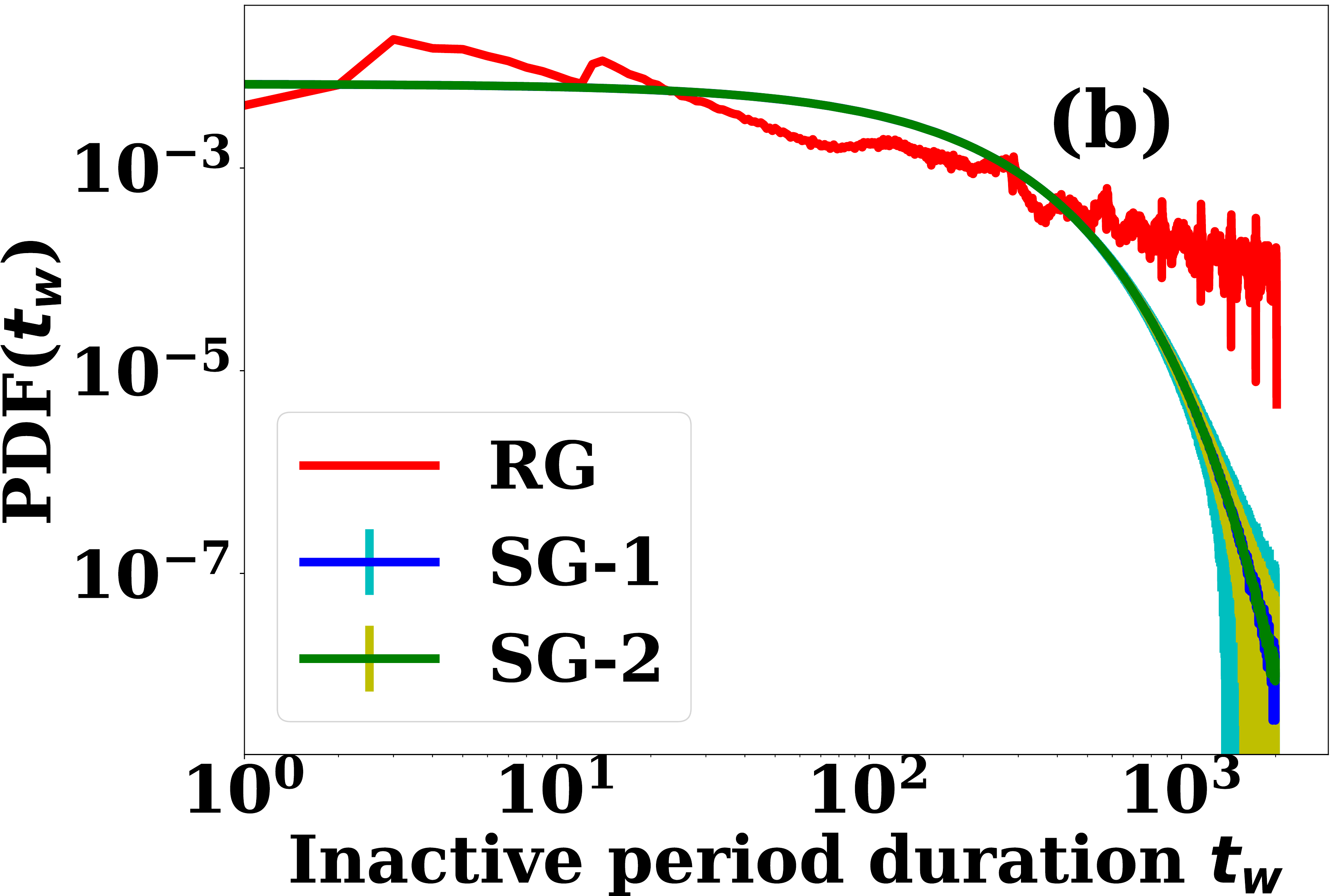}\label{fig:fittb}}\\[-2ex]
     \subfloat{\includegraphics[width=0.49\linewidth, height=2.6cm]{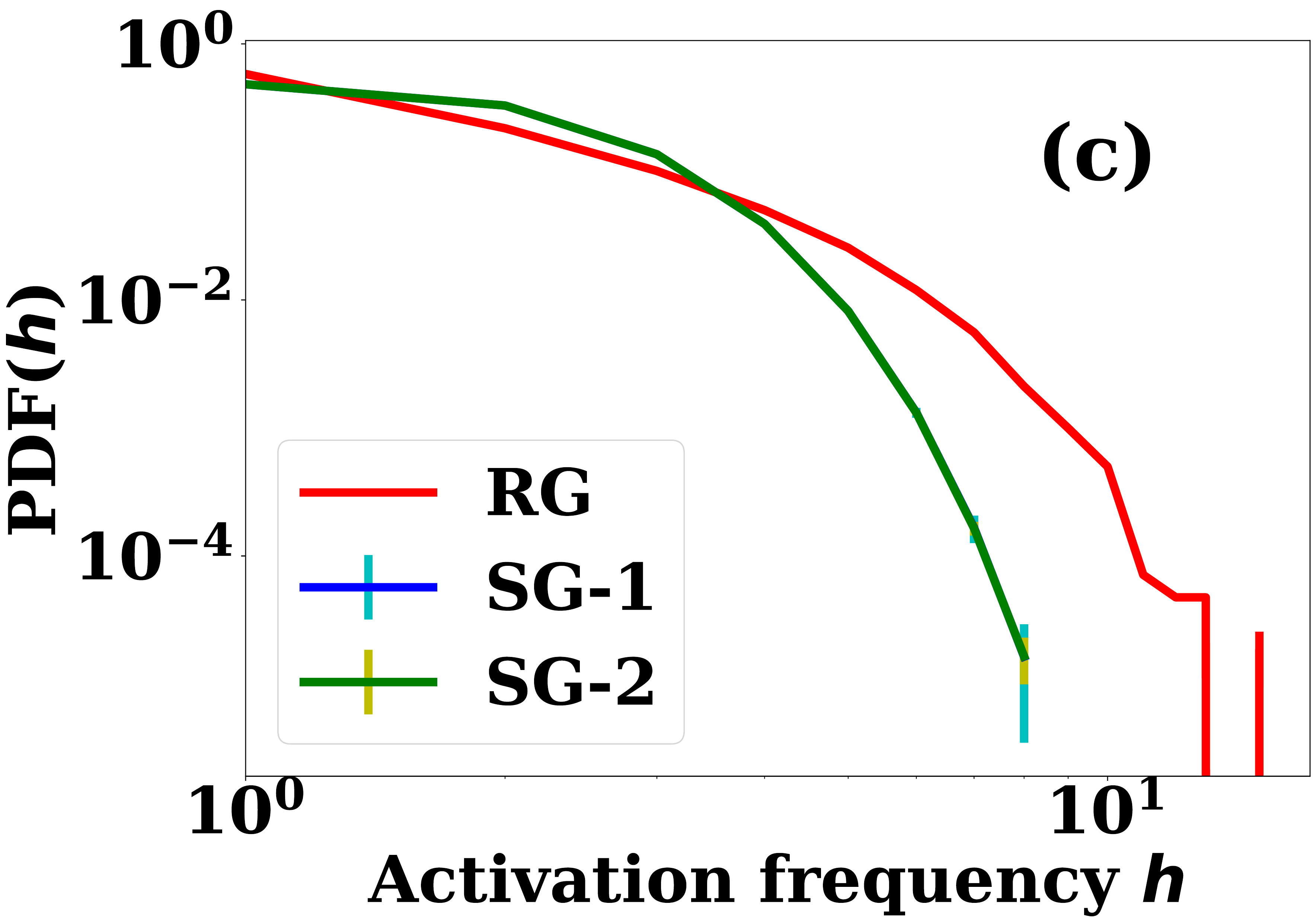}\label{fig:fittc}}~
	 \subfloat{\includegraphics[width=0.49\linewidth, height=2.6cm]{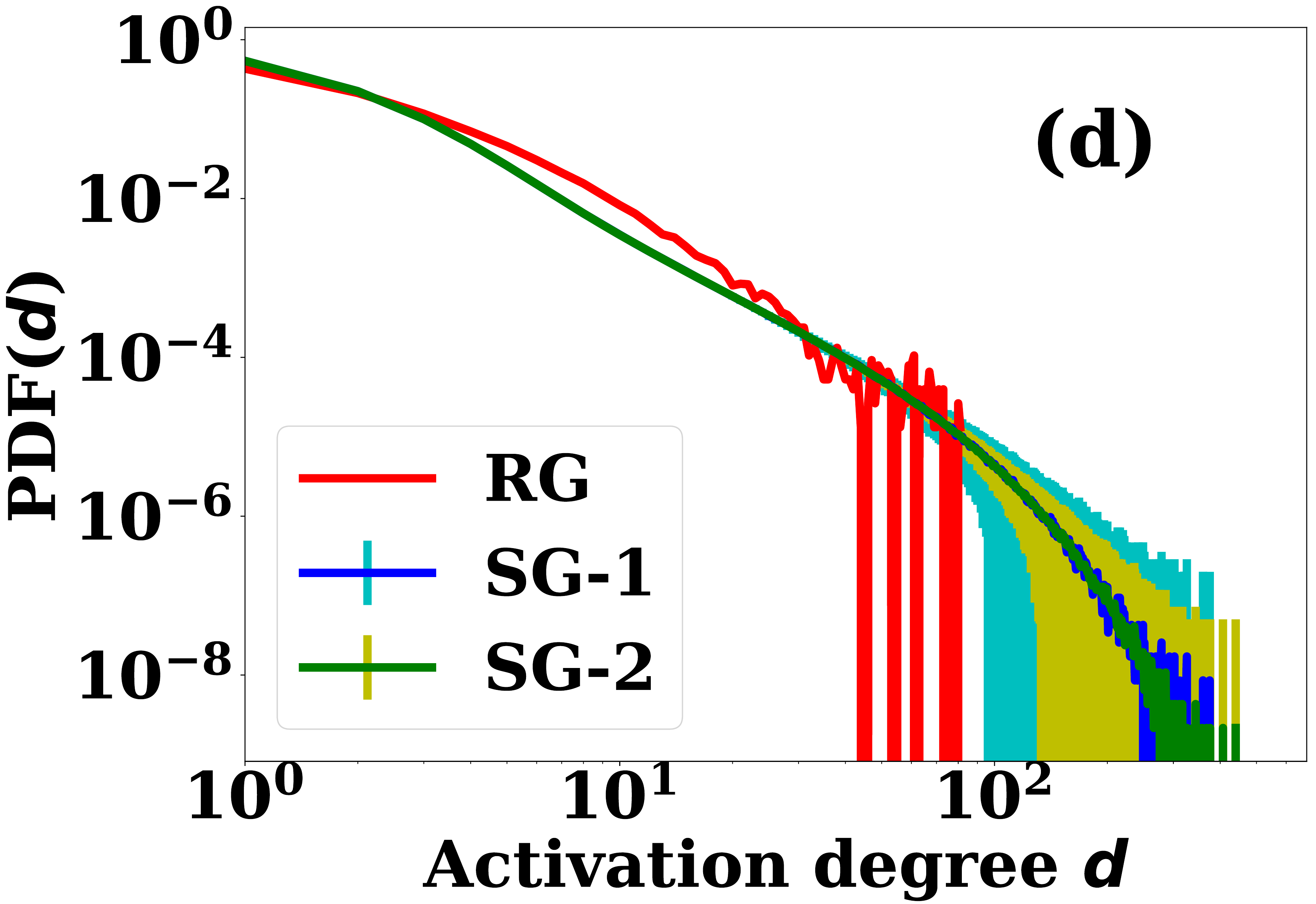}\label{fig:fittd}}\\[-2ex]
     \subfloat{\includegraphics[width=0.49\linewidth, height=2.6cm]{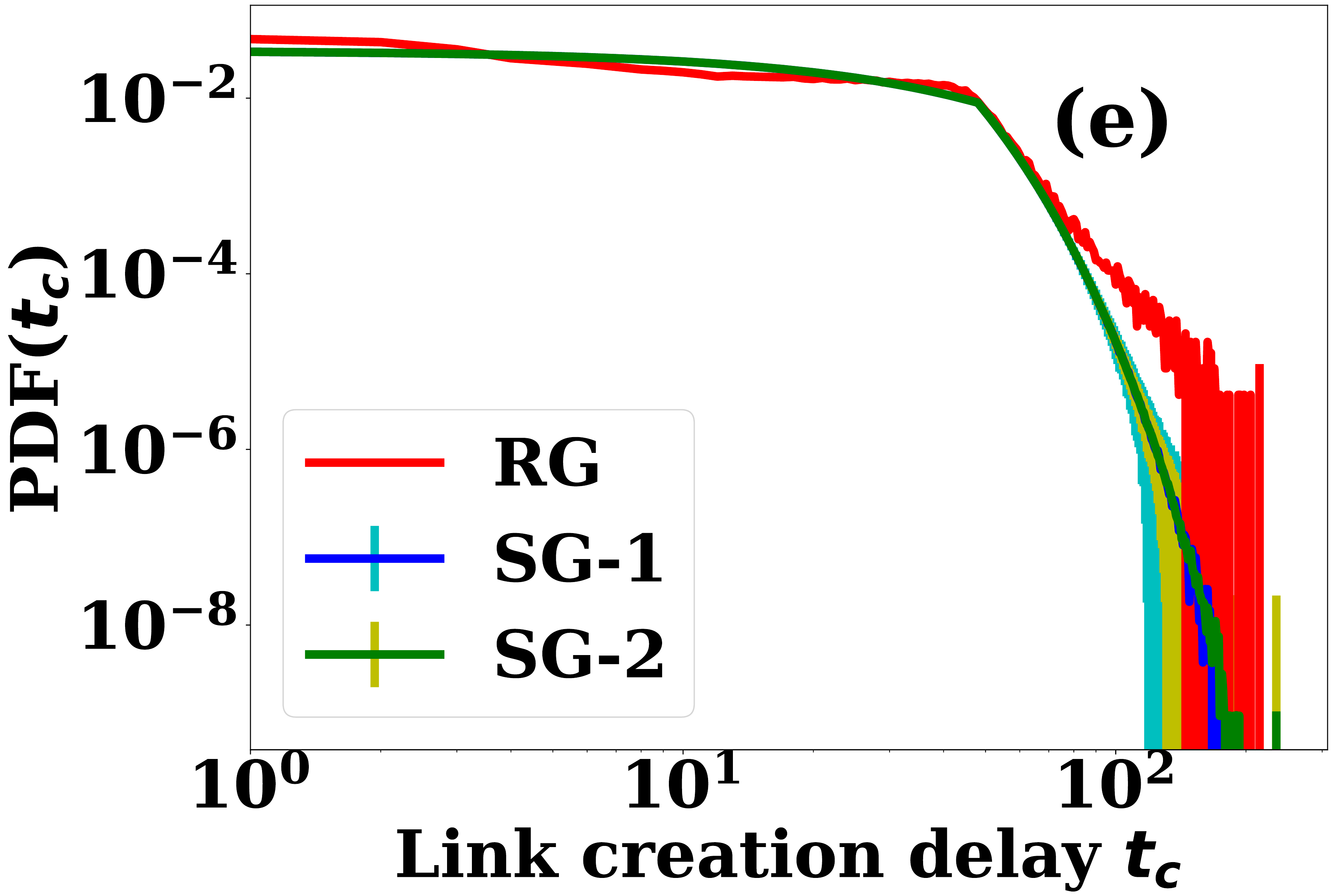}\label{fig:fitte}}~
	 \subfloat{\includegraphics[width=0.49\linewidth, height=2.6cm]{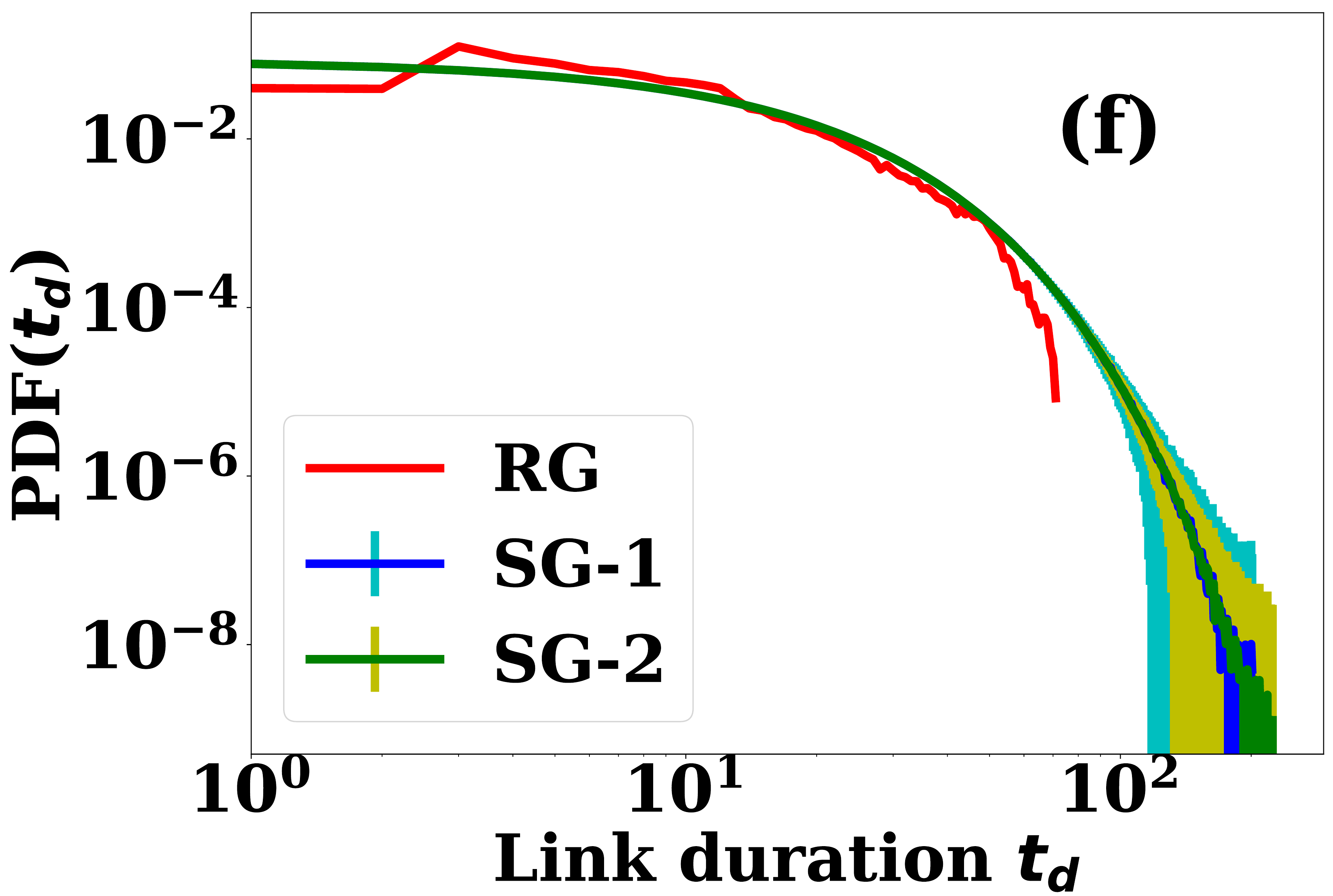}\label{fig:fittf}}\\[-2ex]
     \caption{Comparisons of CIP parameters of real graph (RG) and two generated synthetic SPDT graphs: SG-1 and SG-2} 
     \label{fig:fitt}
     \vspace{-1.3 em}
 \end{figure}

In the previous sections, we have defined five co-located interaction parameters (CIP) namely active period $(t_{a})$, waiting period $(t_{w})$, activation degree $(d)$, link creation delay $(t_{c})$ and link duration $(t_{d})$ that construct the SPDT graph $G=(V,A,E,T)$. For fitting model parameters with real SPDT graph, we extract CIP using updates collected over 07 days form Momo users of Shanghai city. These location updates made by 126K users form 518K active periods which create a SPDT graph of 1.69M SPDT links. We first use Maximum Likelihood Estimation (MLE) techniques and the sample CIP data for finding model parameters. Then, we generate a synthetic SPDT graph (SG-1) of 126K nodes for 7 days using the estimated model parameters and compare the CIP of synthetic graph with the CIP of real graph (RG) made by 7 days updates of 120K users from Beijing city. To understand the model's response across large graph sizes, we also generate another graph (SG-2) of 0.5M nodes for 7 days. The discrete time step of 5 minutes and $\delta =3$ hours are used to generate synthetic graphs. Figure~\ref{fig:fitt} shows results of 500 runs for each graph where periods are in time step of 5 minutes.  

We find the root squared error (RSE) between the generated and real data distributions of CIP parameters as:
\begin{equation}\label{eq:error}
RSE=\sqrt{\sum_{i=1}^{m}(x_i-y_i)^{2}}
\end{equation}
where observed values are grouped in the $m$ bins as they are discrete, $x_i$ is the proportion of observations for the i$^{th}$ bin, $y_i$ is the proportion of empirical dataset values in the i$^{th}$ bin. As RSE are computed from the proportion values, bins are naturally weighted so that bins representing larger proportions of events have higher contributions in error. We plot the distributions of mean values of CIP with deviations of mean in Fig~\ref{fig:fitt}.
 
\subsubsection{\textbf{Node Activation Parameters}}
In our model, active period durations $t_a$ are drawn from a geometric distribution with the scaling parameter $\rho$. The value of $\rho$ can be obtained with sample data using the following MLE condition of geometric distribution:
\begin{equation*}
\hat \rho =\frac{n}{\sum_{k=0}^{n}t_{a}^{k}}
\end{equation*}
where $n$ is the size of sample set $t_a=\{t_a^{1},\ldots,t_a^{n}\}$. We apply $n=518 K$ real active period durations and estimate $\hat \rho=2.83\times 10^{-4}$  s$^{-1}$. The distributions of generated $t_a$ for both graphs SG-1 and SG-2 are shown in Fig~\ref{fig:fitta}. The RSE errors for both SG-1 and SG-2 is 0.065. The model with fitted parameters consistently generates the active period durations $t_a$ for both graph sizes. 

The active period durations $t_a$ have similar patterns for all individuals. However, the inactive period duration of an individual depends on how frequently the individual visits public places. Thus, the distribution of $t_w$ is determined with the distribution of activation frequencies $h$ of individuals. To find $q$, we fit $h$ with the real activation frequencies of momo users. According to Equation~\ref{son}, the probability of transition event $0 \rightarrow 1$ at a time step is: 
\[p_{01}=\frac{\rho q}{q+\rho}\]
Thus, the number of transition events $0 \rightarrow 1$ during $z$ time steps represents the number of activation events $h$. The probability of $h$ activations by the nodes is given by the Binomial distribution as: 
\begin{equation*}
Pr(h\mid q)=\begin{pmatrix}
z\\ 
h
\end{pmatrix}\left(\frac{\rho q}{q+\rho}\right)^{h}\left(1-\frac{\rho q}{q+\rho}\right)^{z-h}
\end{equation*}
The term $\frac{\rho q}{q+\rho}$ becomes small as $\rho=0.085$, $z$ is usually large and $q<1$. Thus, the above equation can be approximated to a Poisson distribution as:
\begin{equation*}
Pr(h\mid q)=\frac{\left(\frac{z \rho q}{q+\rho}\right)^ {h}e^{-\frac{z \rho q}{q+\rho}}}{h!}
\end{equation*}

The MLE condition for the Poisson distribution is given as 
\begin{equation*}
\frac{zq\rho}{q+\rho}=\frac{1}{m}\sum_{i=1}^{m}h_i
\end{equation*}
Using the activation frequencies sample set $h=\{h_1,h_2,\ldots h_m\}$ of size m=126K to MLE equation provides $\hat q=2.23\times 10^{-5}$ s$^{-1}$. The activation frequencies represent the number of active periods a user does in a day. The sample data set is collected over 7 days to find the average values. The generated activation frequencies for SG-1 and SG-2 are presented in Figure~\ref{fig:fittc}, with RSE of 0.077 for both SG-1 and SG-2 compared to real one. We also plot the corresponding waiting periods durations distribution in Figure~\ref{fig:fittb} which follows the distribution of real $t_w$ with RSE around 0.031 in both networks. The $t_w$ is characterized by the irregularity of using Momo Apps.  

\subsubsection{\textbf{Activation Degree Parameters}}
For each active copy, an activation degree $d$ is assigned following Equation~\ref{actdgr}. The value of $d$ depends on the node's public accessibility $\lambda$ drawn from the Power law distribution in Equation~\ref{accesb}. Therefore, the distribution of $d$ in the network will be given for any $\lambda$ as:
\begin{equation*}
Pr(d)=\frac{\beta }{\xi^{\beta}-1}\int_{\xi}^{1}( \lambda^{d-\beta-2} -\lambda^{d-\beta-1}) d\lambda
\end{equation*}
\begin{equation*}
=\frac{\beta}{\xi^{-\beta} - 1}\left(\frac{1-\xi^{d-\beta-1}}{d-\beta-1}-\frac{1-\xi^{d-\beta}}{d-\beta}\right)
\end{equation*}
For estimating the parameters $\hat{\beta}$ and $\hat{\xi}$, we derive the MLE equations and apply the activation degree sample set $d=\{d_{1},\ldots,d_{n}\}$ of size n=518K. We estimate $\hat \beta=2.98$, $\hat \xi=0.25$. We set $\hat \psi =0.999$ as $\lambda$ should be below 1. The generated activation degree distributions for SG-1 and SG-2 with real data presented in Figure~\ref{fig:fittd}. The RSE error is 0.055. The fluctuating error at the tail is due to data sparsity. 

\subsubsection{\textbf{Link Creation Parameters}}
Recall that the maximum link creation delay for a link has the upper bound of the period $t_{a}+\delta$. We apply real active period durations and link creation delays of Momo users to estimate the link creation probability $\hat{p}_{c}$ using the MLE condition:
\begin{equation*}
0=\frac{m}{p_c}-\sum_{k=1}^{l} \frac{\sum_{j=1}^{n}\frac{(t_{c}^{k}-1)(1-(1-p_c)^{t_a^{j}+\delta})+(t_{a}^{j}+\delta)(1-p_c)^{t_a^j +\delta}}{(1-p_c)(1-(1-p_c)^{t^{j}_{a}+\delta})^2}}{\sum_{j=1}^{n} ((1-(1-p_{c})^{t^{j}_{a}+\delta})^{-1}}
\end{equation*}
where $t^{1}_{c},t^{2}_{c},\ldots,t^{l}_{c}$ are sample set of size $l=1.2M$ and $t_a=\{t_a^{1},\ldots,t_a^{n}\}$ with $n=518$K. The estimated value of $\hat p_c$ is 9.33$\times 10^{-5}$ s$^{-1}$. The generated link creation delays are presented in the Figure~\ref{fig:fitte}, where the generated $t_c$ have RSE of 0.035 in comparison with the real distribution. The errors were consistent in both SG-1 and SG-2. Then, we set $p_b=\rho$ for link duration distribution. Figure~\ref{fig:fittf} presents the comparison of generated link durations with real durations which has RSE error of 0.075. The CIP parameters of the generated network are fitted well with the real network parameters made by Momo users. The variations for generated graphs are very low and consistent in both SG-1 and SG-2 networks.

\begin{figure}
\vspace{-1.5 em}
	\subfloat{\includegraphics[width=0.47\linewidth, height=2.6cm]{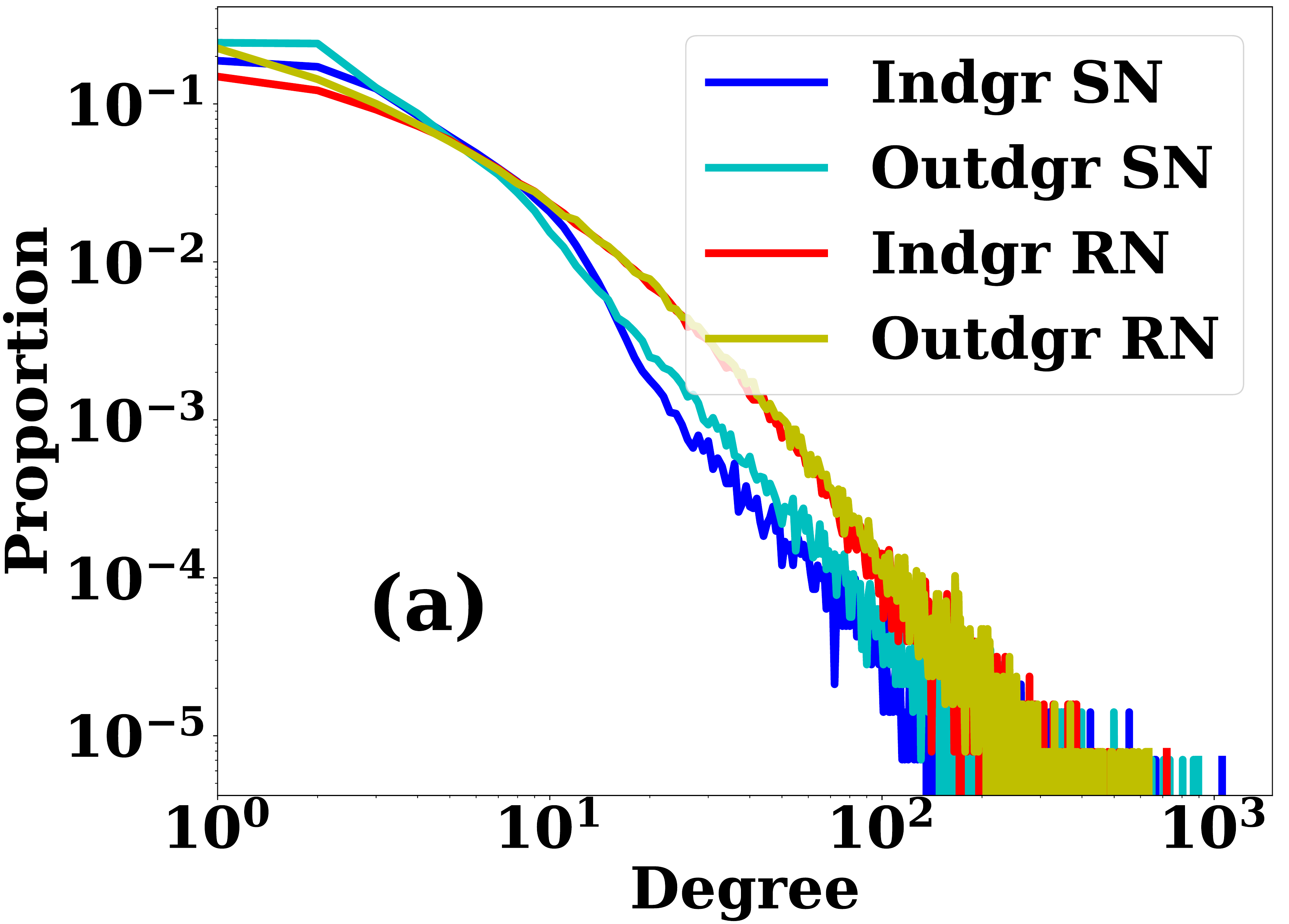}\label{fig:statdg}}~
	\subfloat{\includegraphics[width=0.47\linewidth, height=2.6cm]{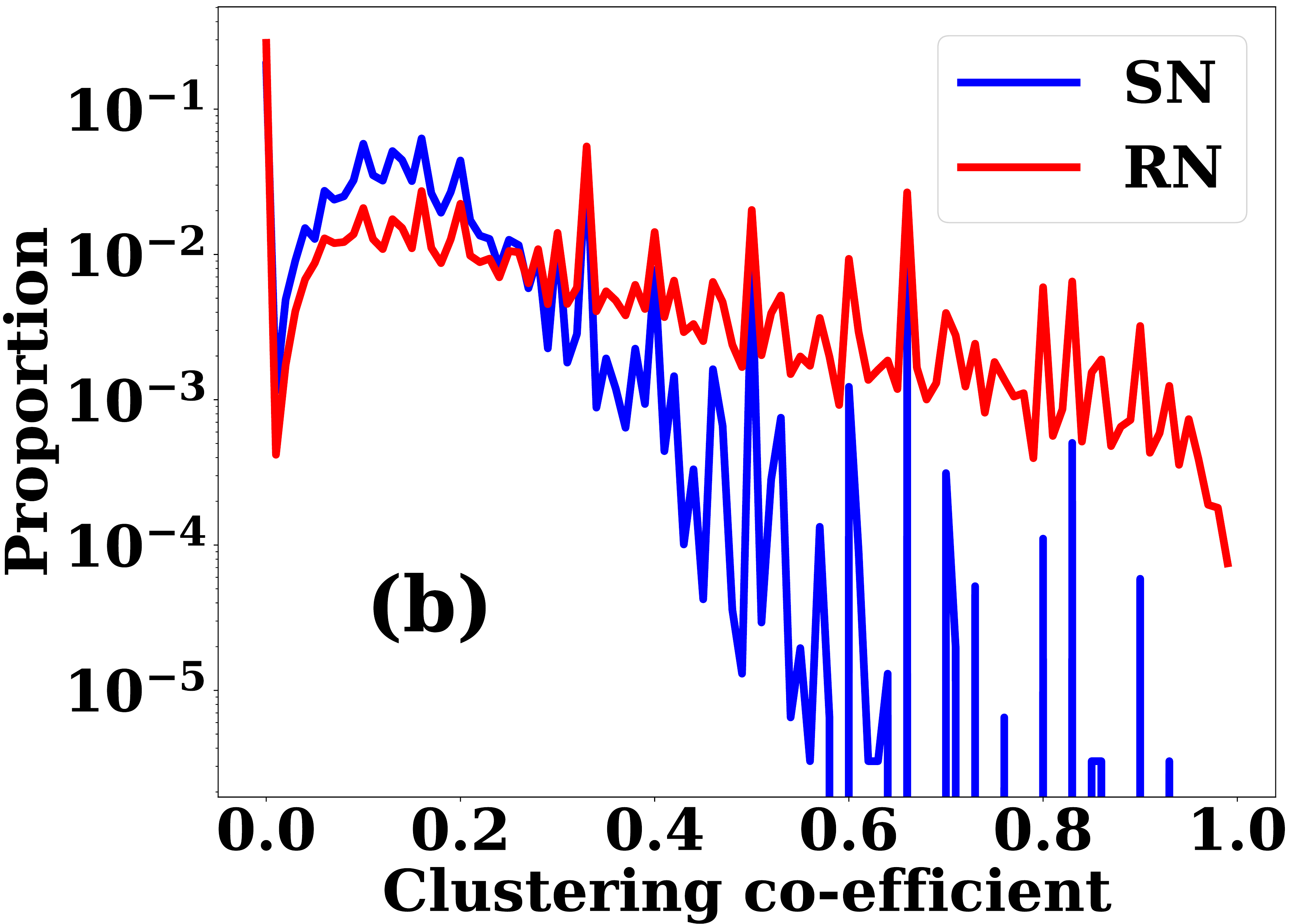}\label{fig:statcf}}
    \vspace{-1.3 em}
      \caption{Static properties of graphs: a) degree centralities and b) clustering co-efficients}
      \label{fig:staticpro}
      \vspace{-1.2 em}
\end{figure}
\subsection{Network Properties}
The previous section has addressed the temporal aspects of interactions under SPDT graph fitting CIP parameters to empirical data. This section explores the fitted model's ability to reproduce important static graph properties of empirical networks. The model's parameters have been tuned using the updates from Shanghai. We now utilize the location updates from Beijing to compare the network properties. We generate a synthetic SPDT graph of two weeks with 147K nodes for comparison against an empirical dataset from Momo with the same number of users and duration. We summarize the generated graph by a static graph where a directed edge between two nodes is created if they have at least one SPDT link from host node to neighbor node at any time. We first analyze the degree centrality that quantifies the extent of a node's connectedness to other nodes~\cite{freeman1978centrality}. In a disease spread context, nodes with higher degree centrality get infected quickly as well as infect a higher number of other nodes~\cite{de2014role}. In our model, the growth of the contact set of node $i$ is determined by $\lambda_{i}$ and the neighbor selection process defined by $p(n_t+1)=n_t/(n_t+\eta)$. The value of $\eta$ controls the degree to which nodes expand their contact set size. We select $\eta=1$ to provide reasonable growth in contact set sizes through the influence of $\lambda$ which varies across nodes, while the selection of an optimal value of $\eta$ is beyond the scope of this paper. Another desirable feature is that nodes which have more directed links to other nodes also receive more links. We explore this effect in Figure~\ref{fig:statdg}, which plots the distribution of in-degree and out-degree of nodes for both real (RN) and synthetic graphs(SN). The degree distributions are similar in both networks. The correlation between the in-degree and out-degree are about $0.895$ for both graphs.

While degree centrality highlights the node connectivity, we use the local clustering coefficient to study the social structure of the network to understand the community structure in the generated graphs~\cite{laurent2015calls}. A node selects a new neighbor from its second degree neighbor set, i.e. it's current neighbor's neighbors. For analysis, we convert the directed SPDT links into undirected links
and compute the clustering coefficients for each node. We present the results in Figure~\ref{fig:statcf}. The average clustering coefficient in the real graph is $0.11$ while the synthetic graph has $0.08$. The RSE error between the distributions of clustering co-efficient is $0.0623$. We attribute the difference between the two graphs to the distinctions between the randomized links in the synthetic graph and the well-known non-random network structure of empirical %social networks~\cite{}, which we leave for future
social networks, which we leave for future work. Still, our results reflect that the proposed graph model can approach the empirical social structure even with simple methods of neighbor selection.

%% file: diffussion.tex
\begin{figure}
\vspace{-1.7 em}
	\subfloat{\includegraphics[width=0.48\linewidth, height=2.8 cm]{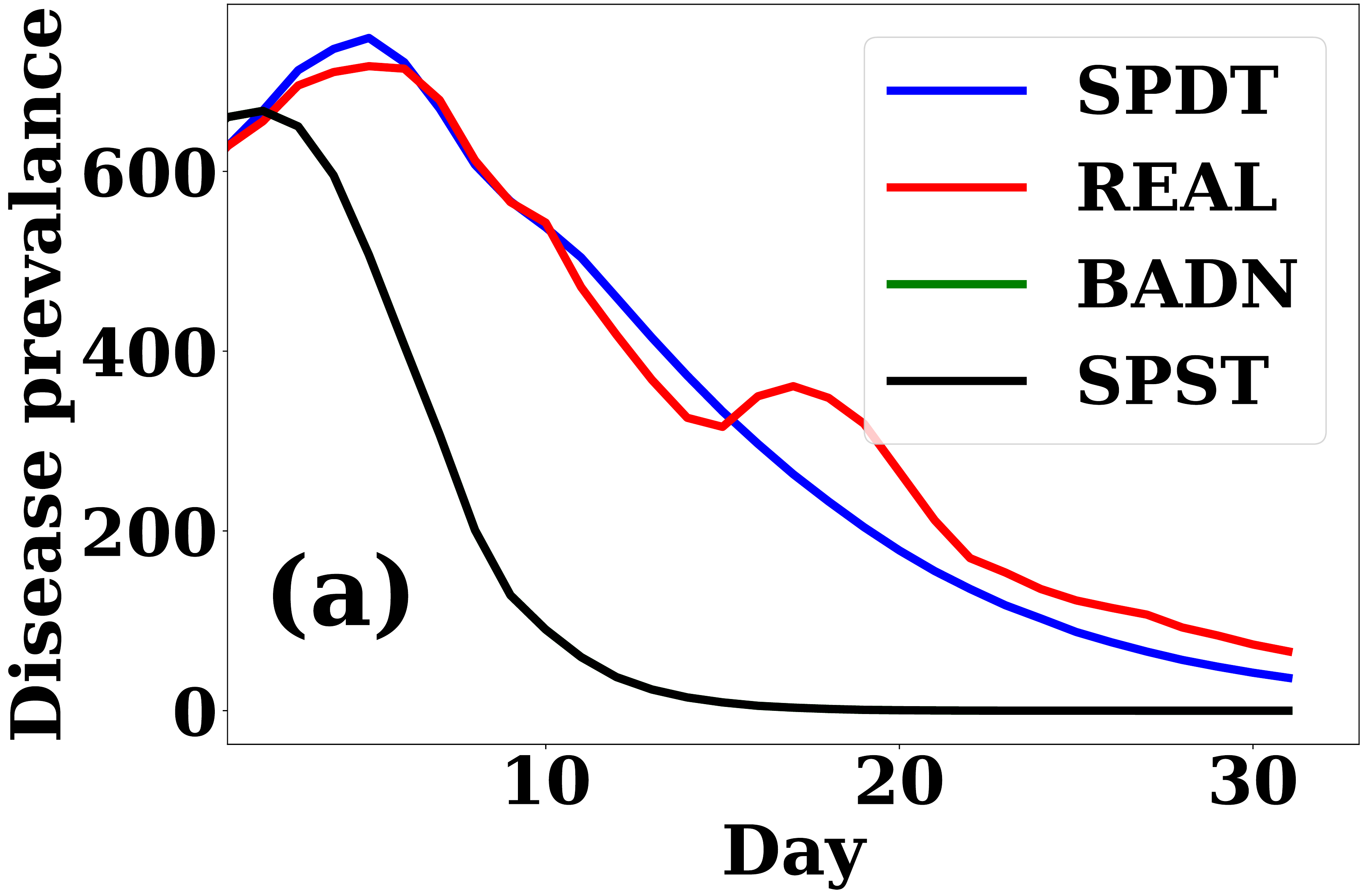}}~
    \subfloat{\includegraphics[width=0.48\linewidth, height=2.8 cm]{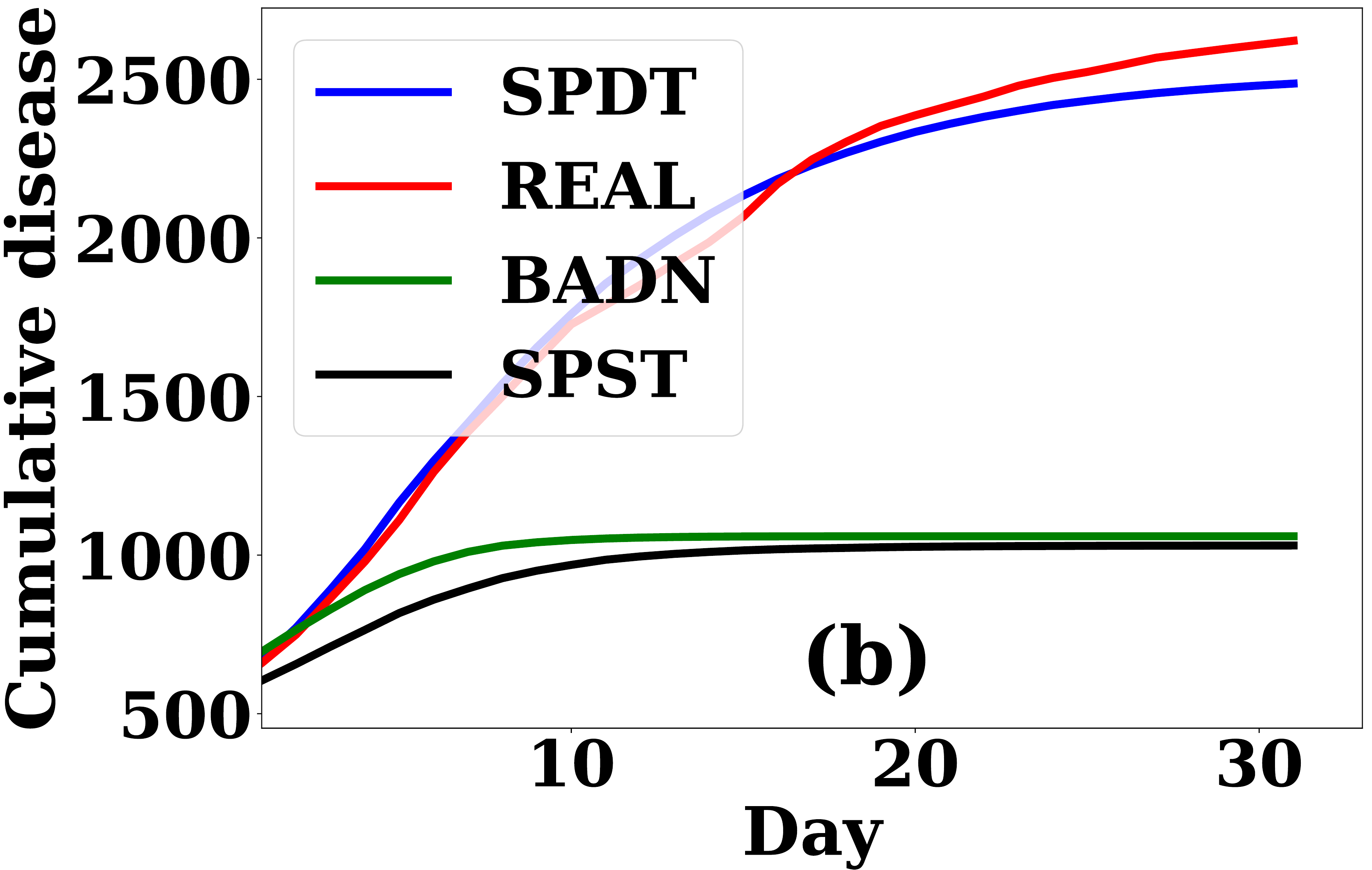}}\\[-2ex]
     \subfloat{\includegraphics[width=0.48\linewidth, height=2.8 cm]{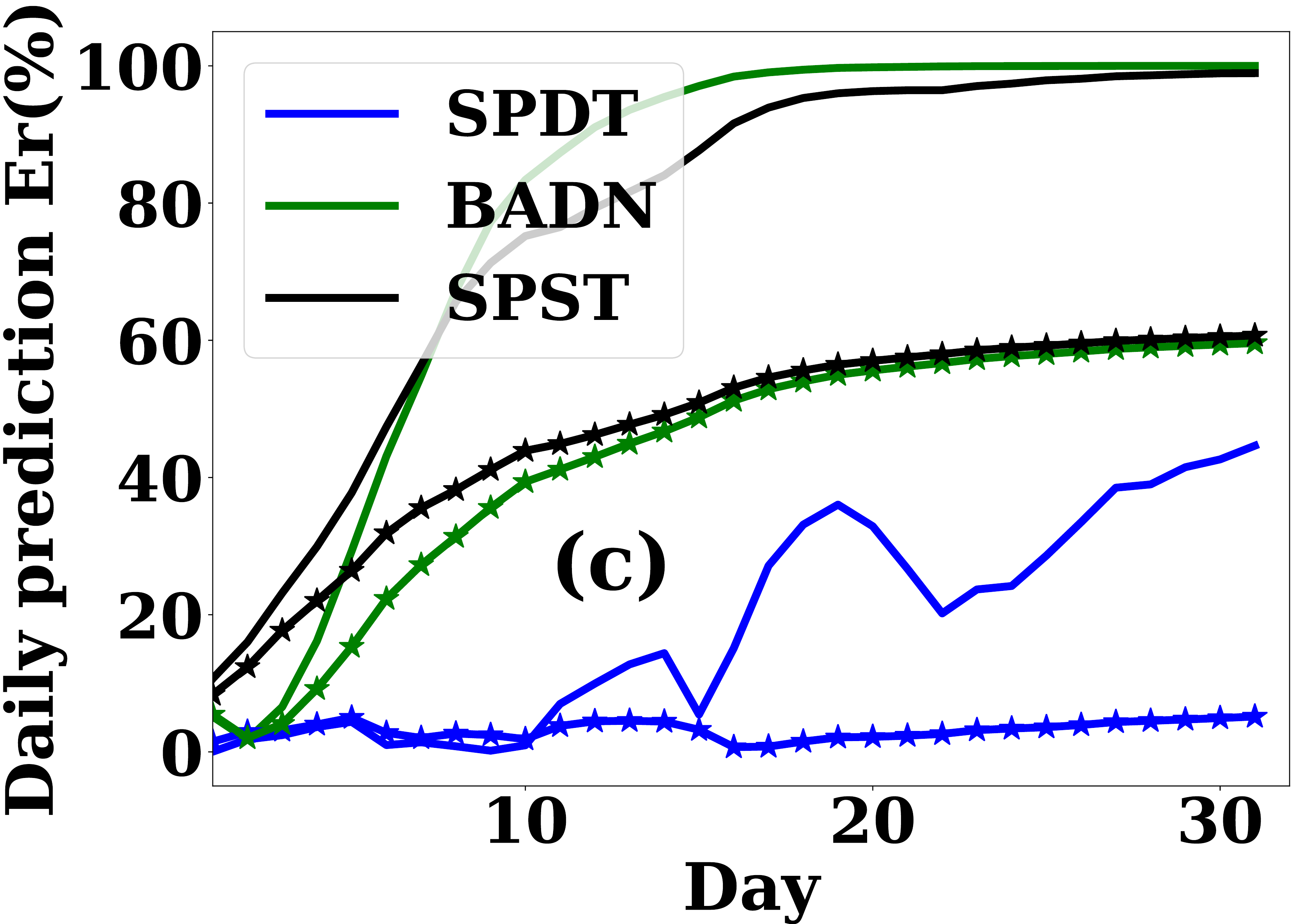}}~
     \subfloat{\includegraphics[width=0.48\linewidth, height=2.8cm]{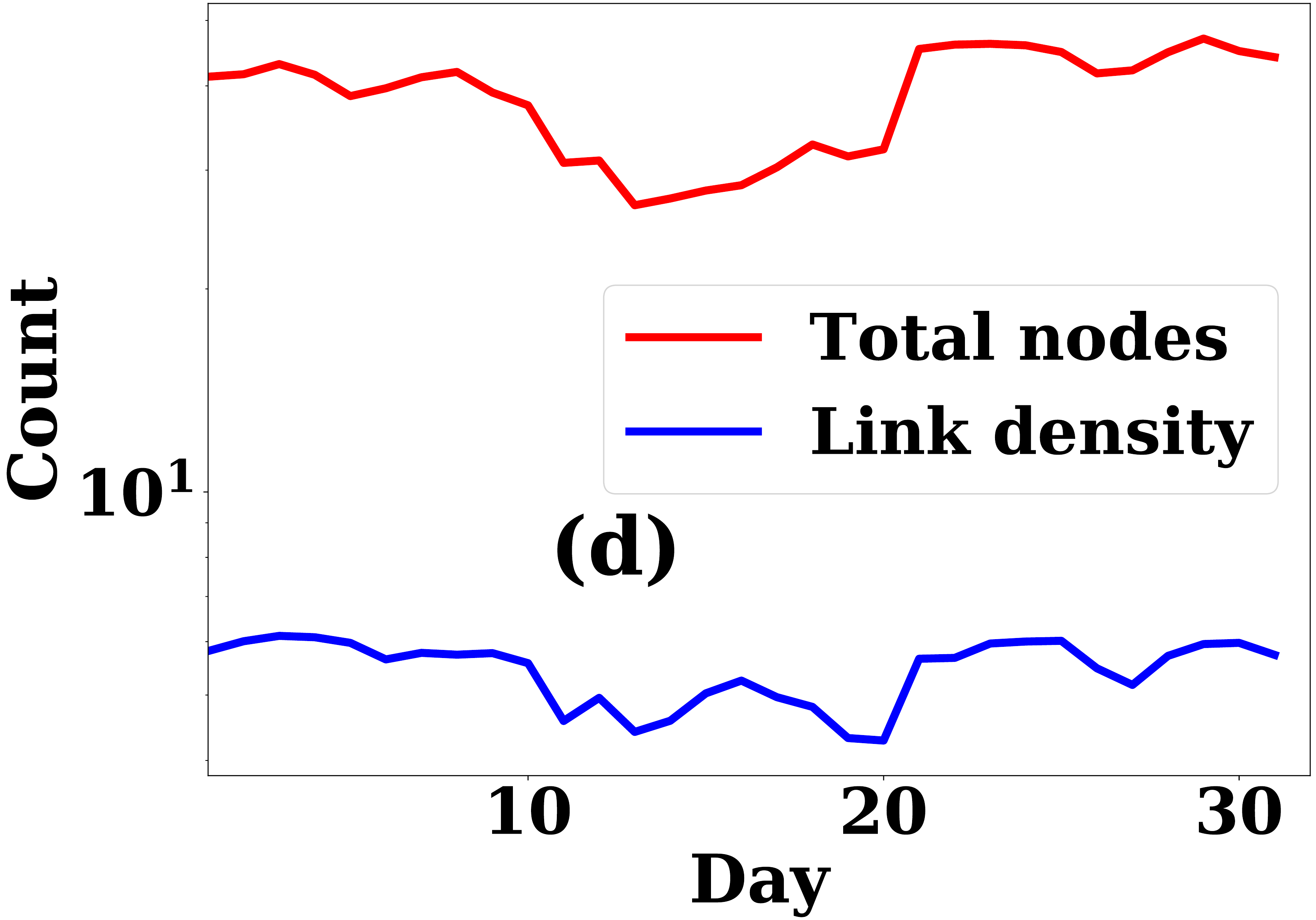}}\\[-2ex]
     \subfloat{\includegraphics[width=0.48\linewidth, height=2.8cm]{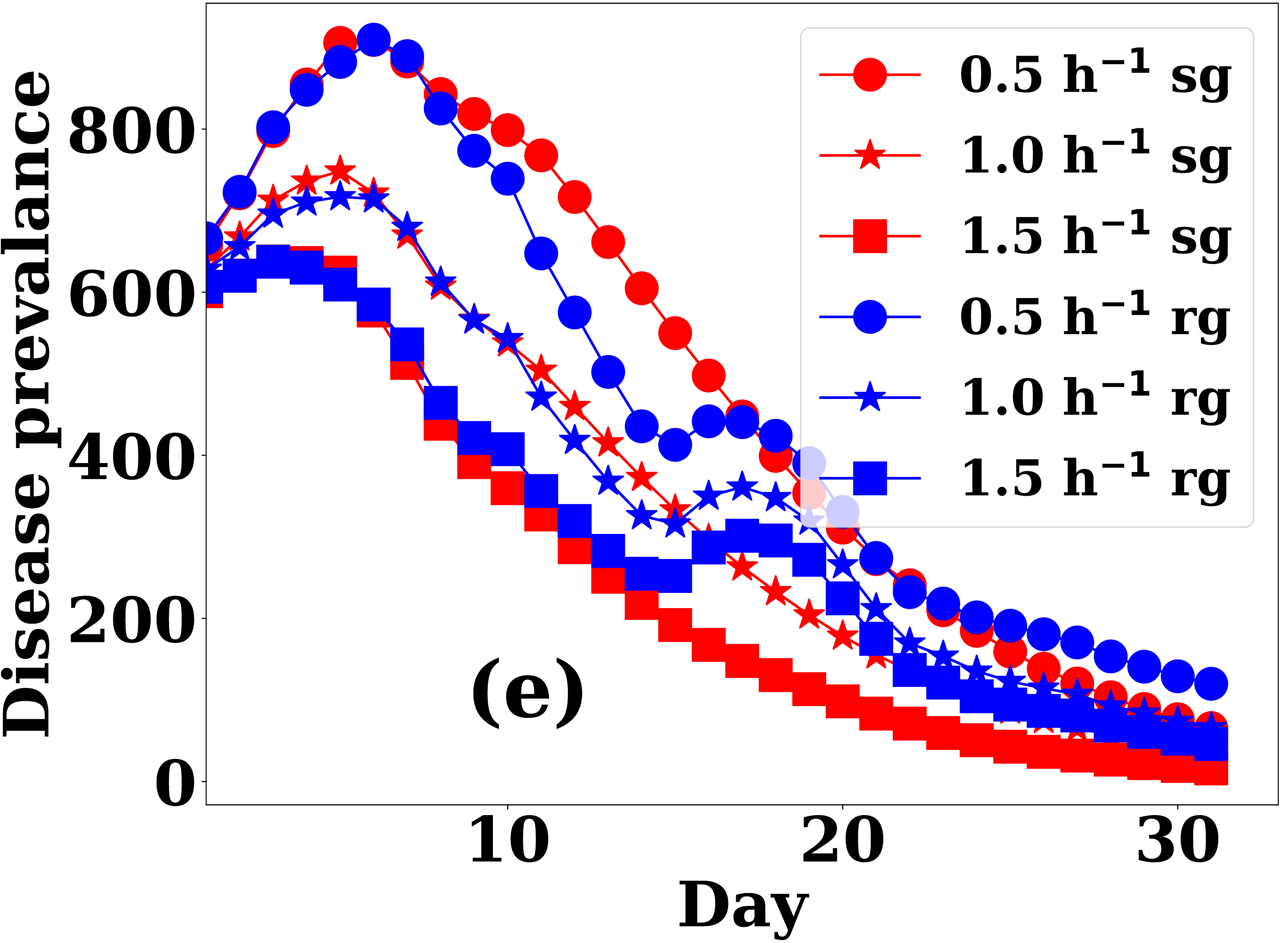}}~
     \subfloat{\includegraphics[width=0.48\linewidth, height=2.8cm]{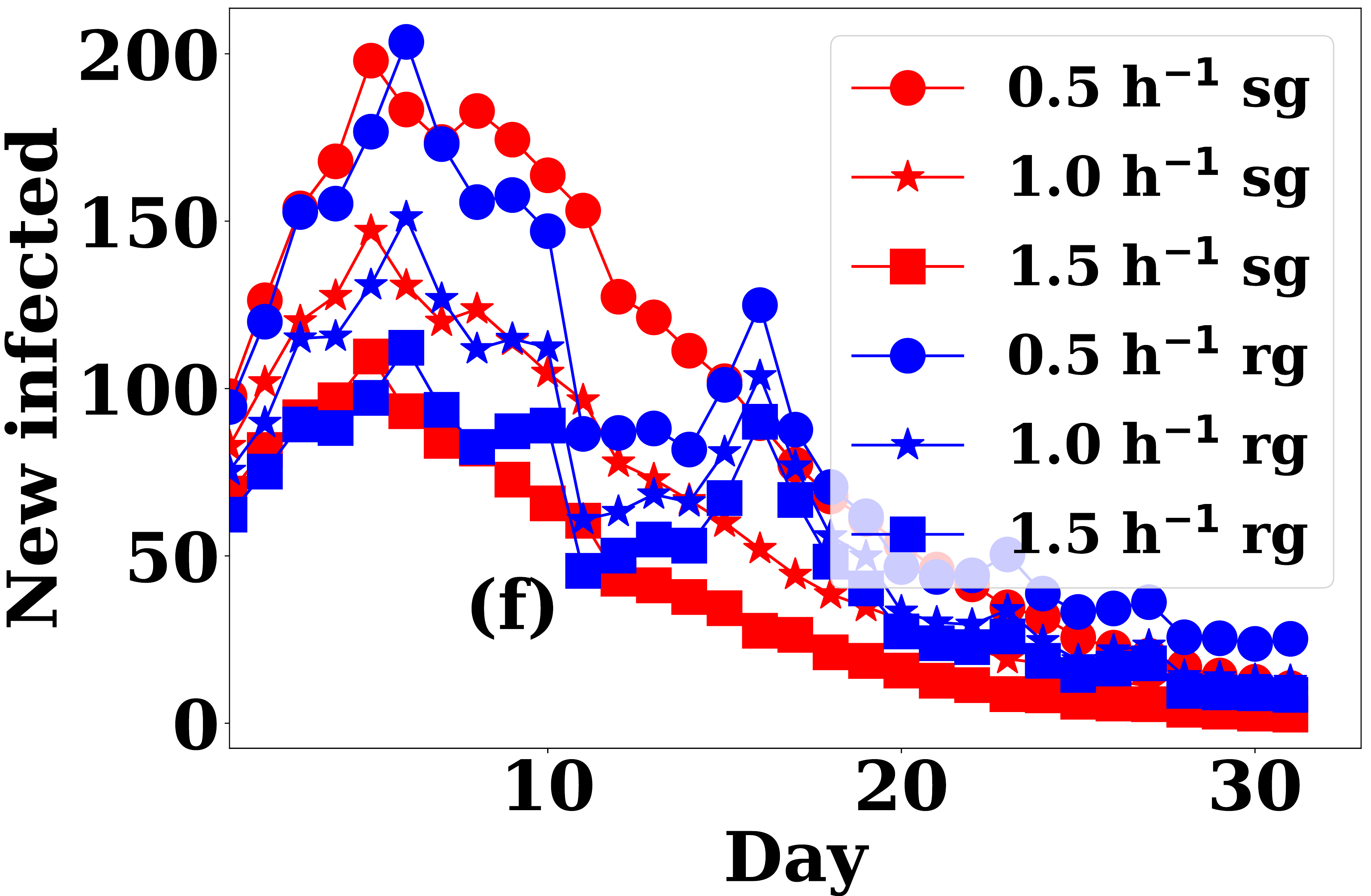}}
     \vspace{-1.0 em}
     \caption{Diffusion dynamics on the sparse graph: (a) disease prevalences $I_p$ for real graph, SPDT model, BADN graph and SPST graph, (b) final epidemic size, (c) prediction error for $I_p$: * lines for cumulative and other for daily (d) number of Momo users (K) and link densities per user, (e) $I_p$ for various $r$ in RG and SG, and (f) new infections for various $r$} 
     \label{fig:sdif}
     \vspace{-1.6 em}
\end{figure}

\begin{figure*}
\vspace{-1.5 em}
	\subfloat{\includegraphics[width=0.19\textwidth, height=2.8 cm]{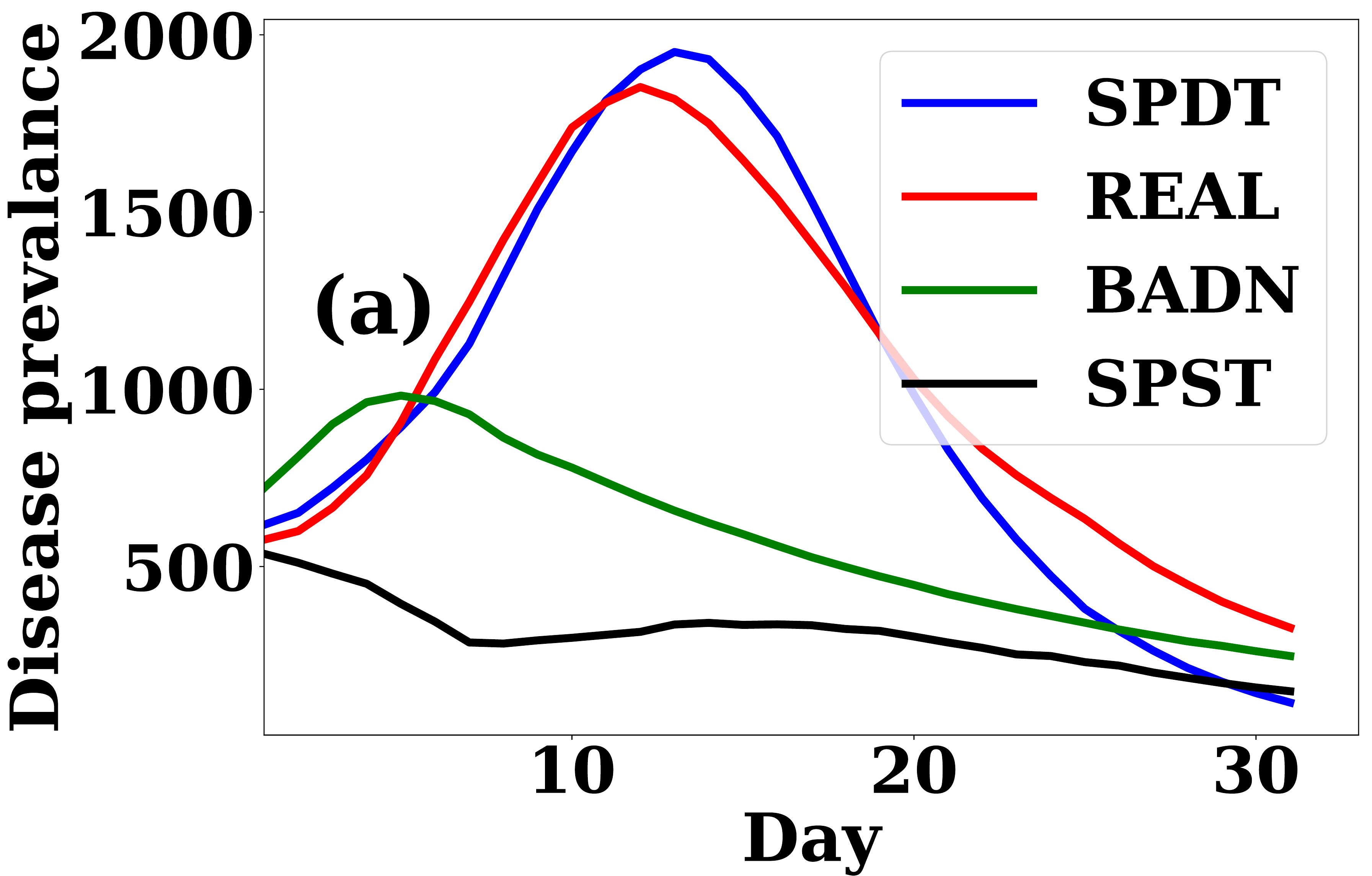}}~
    \subfloat{\includegraphics[width=0.19\textwidth, height=2.8 cm]{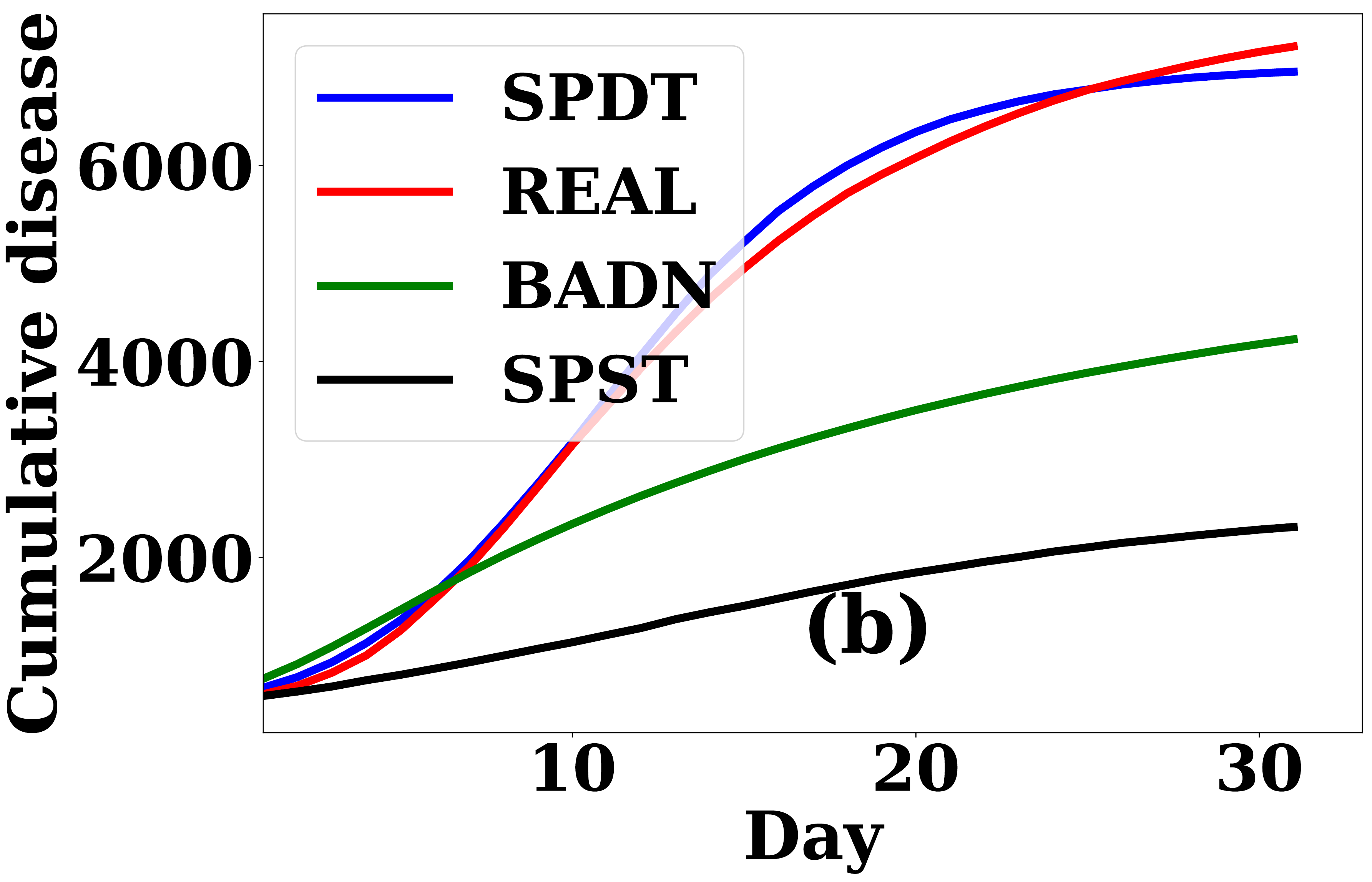}}~
     \subfloat{\includegraphics[width=0.18\textwidth, height=2.9 cm]{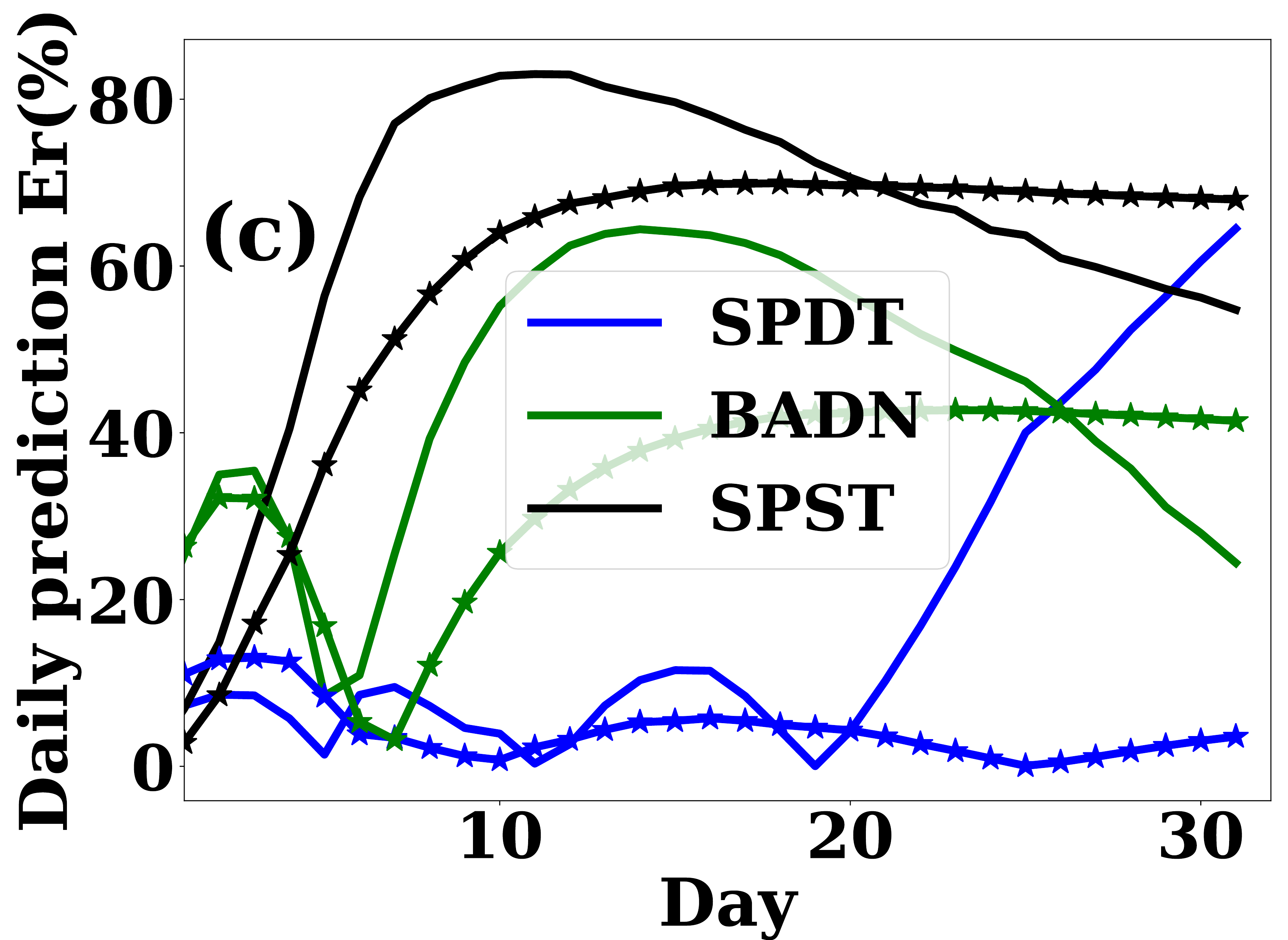}}~
     \subfloat{\includegraphics[width=0.22\textwidth, height=2.8cm]{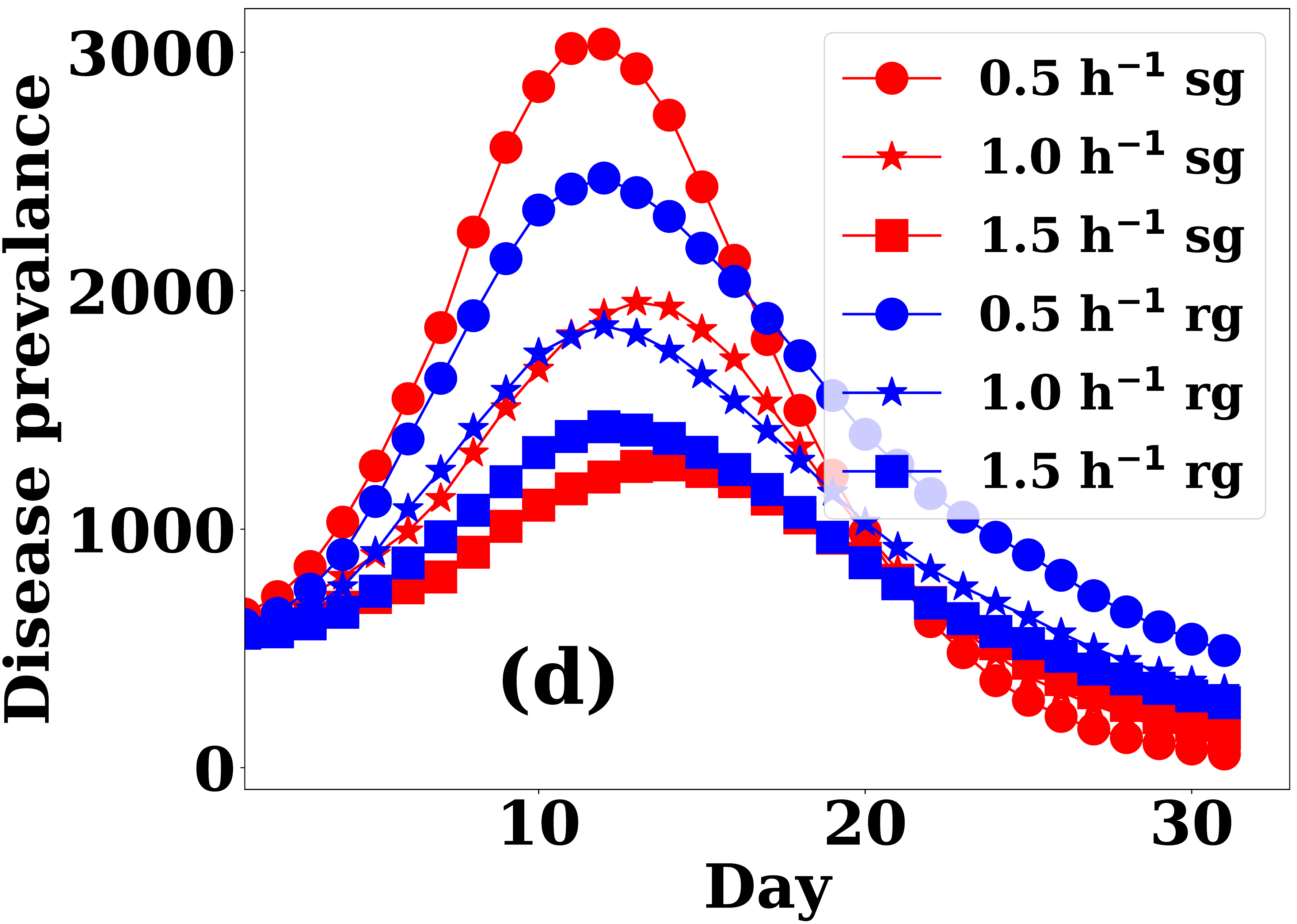}}~
     \subfloat{\includegraphics[width=0.22\textwidth, height=2.8cm]{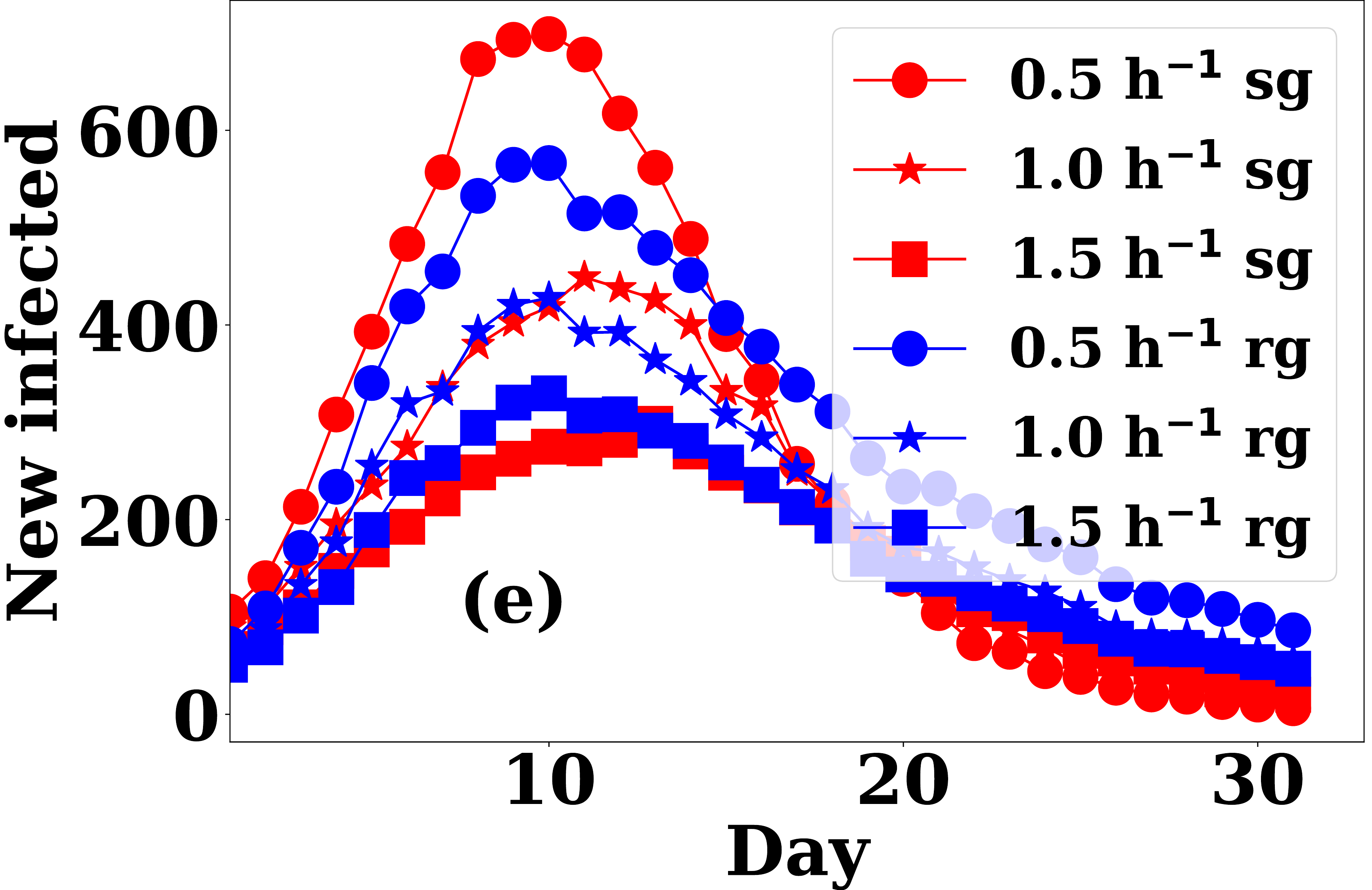}}~
%      \subfloat{\includegraphics[width=0.2\textwidth, height=2.8cm]{cnt_small.pdf}}
     \vspace{-1.0 em}
     \caption{Diffusion dynamics on the dense graphs: (a) disease prevalences $I_p$ for real graph, SPDT model, BADN graph and SPST graph, (b) final epidemic sizes (c) prediction errors, (d) $I_p$ for various r in RG and SG, and (e) new infections for various r} 
     \label{fig:sdifl}
     \vspace{-1.3 em}
\end{figure*}

\section{Model Validation}
In this section, we validate the proposed model simulating SPDT process on the generated synthetic graph and real contact graph. Accordingly, airborne disease spreading is simulated on the synthetic and real traces. The simulations are also conducted on another synthetic graph constructed according to the basic activity driven networks (BADN) model~\cite{perra2012activity} to understand how well the proposed model capture diffusion dynamics comparing to the current graph models. In our simulations, the infection probability $P_I$ for inhaling $E_T$ dose of infectious particles by a node is 
\begin{equation}\label{expos}
P_I=1-e^{-\sigma E_T}
\end{equation}
where $\sigma$ is the infectiousness of particles~\cite{shahzamal2017airborne}. The value of E for a SPDT link $e_{vu}=\left(v(t_s,t_l),u,t_s^{\prime},t_l^{\prime}\right)$ can be calculated as: 
\begin{equation*}
E=\frac{gp}{Vr^2}\left[r\left(t_i-t_s^{\prime}\right)+ e^{rt_{l}}\left(e^{-rt_i}-e^{-rt_l^{\prime}} \right)+e^{rt_{s}}\left(e^{-rt_l^{\prime}}-e^{-rt_s^{\prime}} \right)\right]
\end{equation*}
g is the particle generation rate of an infected individual, p is the pulmonary rate of a susceptible individual, V is the volume of the interaction proximity, r is the particle removal rate from the proximity and $t_i$ is the time depends on the link components: $t_i=t_l^{\prime}$ if link has only direct component, $t_i=t_s^{\prime}$ if link has only indirect component and $t_i=t_l$ for both components. In simulations, disease propagates according to the Susceptible-Infected-Recovered (SIR) epidemic model. A susceptible node's daily exposure for all links from infected nodes is calculated and its infection probability $P_{I}$ is determined. Based on this probability, the node's status is updated at one day intervals. Once infected, a node continues infecting others for a number of days selected uniformly from range of 3 to 5 days, after  the node  recovers~\cite{spricer2017characterizing}. For simulation, $r$ is taken randomly in the range $[0.25,8] h^{-1}$ around a median, $g=0.304 PFU/s$, pulmonary rate $q=7.5 L/min$ and $V=2512 M^{3}$ assuming 20m radius and 2m height of interaction area~\cite{shi2015air,shahzamal2017airborne}. The required exposure to induce disease for 50$\%$ susceptible individual is found in the range 0.69 to 3.5 PFU. If we take the mid range value of PFU for causing 50$\%$ infection, the value of $\sigma$ will be 0.33~\cite{alford1966human}. All simulations start by randomly selecting 500 seed nodes and continues for 32 days.

\begin{table}
\vspace{-0.0 em}
\caption{Summary of simulation results}
\vspace{-1.0 em}
\label{tab.2}
\resizebox{0.48\textwidth}{!}{
\begin{tabular}{|cccccccc|}
\hline
\textbf{Graphs} & \textbf{r$h^{-1}$} & \textbf{Peak $I_p$ }& \textbf{Peak Day} & \textbf{Total I} & \textbf{Total Er($\%$)} & \textbf{Mean Er($\%$)} & \textbf{STD Er($\%$)}\\
\hline
\multicolumn{8}{|c|}{Sparse graphs}\\
\hline
BADN & 1.0 & 631 & 3 & 1069 & 67 & 78 & 33\\
SPST & 1.0 & 568 & 2 & 1030 & 68 & 78 & 27\\
\hline
 & 0.5 & 909 & 7 & 3462 & 3 & 14 & 15\\
SPDT & 1.0 & 748 & 6 & 2486 & 5 & 17 & 15\\
& 1.5 & 637 &4 & 1818 & 16 & 31 & 26\\
\hline
 &  0.5 & 909 & 7 & 3355 & & &\\
Real & 1.0 & 717 & 6 & 2621 & & &\\
 & 1.5 & 637 & 4 & 2171 & & &\\
\hline
\multicolumn{8}{|c|}{Dense graphs}\\
\hline
BADN & 1.0 & 1057 & 6 & 4225 & 42 & 43 & 17\\
SPST & 1.0 & 543 & 2 & 2310 & 68 & 62  & 22\\
\hline
 & 0.5 & 3034 & 13 & 9143 & 4 & 34 & 27\\
SPDT & 1.0 & 1951 & 14 & 6956 & 4 & 18 & 19\\
& 1.5 & 1269  & 15 & 5427 & 7 & 12 & 12\\
\hline
 &  0.5 & 2472 & 13 & 9544 & & &\\
Real & 1.0 & 1852 & 13 & 7313 & & &\\
 & 1.5 & 1429 & 13 & 5862 & & &\\
\hline
\end{tabular}
}
\vspace{-1.4 em}
\end{table}

As  SPDT diffusion changes with $r$, we verify how well the model captures the changes in diffusion dynamics due to changes in $r$. We run simulations on various graphs for values of 0.5,1, and 1.5 $h^{-1}$. The summary of the results is presented in the Table~\ref{tab.2}. First, we chose the real SPDT graphs made by Momo users of Beijing. This graph has total 297K users that create 6.9 million links through 2.2M active periods. We generate a similar SPDT graph with the fitted model parameters. We run 500 simulations for both graphs choosing the median $1.0 h^{-1}$ of r and disease prevalences $I_p$ over simulation days are presented in Fig~\ref{fig:sdif}a. The prevalence $I_p$ shows closely matching trends between the real and synthetic graphs with a concurrent peak at day 7. However, the intensities of $I_p$ vary for some days after day 15. The variations of $I_p$ for the real graph arise from fluctuations in the number of users and link densities~(see Fig~\ref{fig:sdif}d), whereas the SPDT model assumes a constant number of users in the network. Having increased daily prediction error, our model still maintains a cumulative prediction error around 5\% as the new infection rates are relatively small after day 10.

We also compare the model performance simulating diffusion on  a BADN graph where nodes activate at each time step with a probability $b$ and generate $m$ links to others nodes. Analysis of Momo users reveal that their average stay periods $\Delta t=1/\rho$ are 50 minutes and on average create $m=2$ links during an activation. Thus, we define the activation potential for a node as $p=f \Delta t/T$, where $f$ is the activation frequency=3 per day, and generate a BADN graph for 297K nodes. We run simulations for 500 times and results are presented in Fig~\ref{fig:sdif}a. It has infected 1069 nodes in total which is one third of the real graph. Then, the simulation is run on the SPST graph which is obtained by removing indirect paths from SPDT model. This provides similar output of BADN. We calculate the absolute percentage error (APE) for infection events as: 
\[ APE=100\times \frac{I_r-I_o}{I_r}\]
% \begin{equation*}
% APE=100\times \frac{I_r-I_o}{I_r}
% \end{equation*}
where $I_r$ is the number of infection events in the real network and $I_o$ is the infection event in the corresponding observed graph. We calculate the mean APE (MAPE) for the disease prevalence and APE for cumulative disease cases for all graphs. The daily prediction errors in APE of disease prevalence for different models are shown in Fig~\ref{fig:sdif}(c). BADN and SPST have high MAPE error $78\%$ with standard deviation $33\%$ while SPDT graph shows $17\%$ with standard deviation of $15\%$ at particle removal rate $r=1h^{-1}$, which reduces the daily prediction error by nearly 82\%. The prevalence trends for BADN and SPST do not capture the real dynamics, confirming the superiority of the proposed SPDT graph model.

We next evaluate the model's sensitivity to diffusion parameters. We simulate disease diffusion for three different values of $r=\{0.5,1.0,1.5\} h^{-1}$ on the real graphs (RG) and SPDT graph (SG) while other parameters are kept the same. Results for 500 simulations are presented in Fig~\ref{fig:sdif}e for disease prevalence and Fig~\ref{fig:sdif}f is for new daily infections. Similar to Fig~\ref{fig:sdif}a, link density variations impact the intensities of $I_p$ in the real graph. However, at higher values of $r$ the difference shrinks. This is because both graphs approach SPST as $r$ increases, which reduces the indirect transmission period. The results for $r=1.5 h^{-1}$ is more representative for such situations. The prevalence $I_p$ increases during the later days in the real network as many neighbors are infected early in the SPDT graph, reducing new infection rates at the tail. The similar situation is found in Fig~\ref{fig:sdif}(d). The new infection rate in Fig~\ref{fig:sdif}f is more random for the real network and strongly follows the link densities.

To understand the response of the graph model for larger scale simulation with higherlink densities, we reconstruct a large and dense graph from the real graph dynamics. In the reconstructed graph, we populate any days during which a user has no Momo updates with contacts selected from other days during which the user does have updates. We randomly copy a day from the users' available days to an absence day without changing link properties~\cite{machens2013infectious}. We consider a corresponding synthetic SPDT graph of 297K nodes and a BADN graph fitting with the reconstructed graph. The simulation results presented in Figure~\ref{fig:sdifl} show that the SPDT graph reduces the daily prediction error by 58\% and 71\% compared to BADN and SPST respectively, and it reduces the cumulative prediction error by 90\% and 95\% to BADN and SPST graphs respectively. The daily prediction error of SPDT graph grows sharply after day 20 compared to the real graph, with SPDT graph underestimating the number of cases. This deviation can be explained as follows. When the real SPDT graph is reconstructed copying available day links to missing, same links of the nodes who have links for only one day is copied to other 31 days. If one of these nodes is infected, they transmit disease to the same neighbor nodes for whole infectious period. In the proposed model, however, nodes always have some probability to connect to a new neighbor and hence the the contact set size grows, which is a higher chance to cause more infections. Thus, new infection rate in the SPDT graph model before touching peak is higher than the infection rate in real graph. As susceptible reduces quickly in SPDT model at early days of simulation, new infection rate falls faster in the later days. The MAPE error reduces from $34\%$ to $12\%$ as $r$ increases. This is because the indirect paths dominate more at lower $r$ causing more new infections. Still, our model maintains a cumulative prediction error of 4-7\% across all configurations, further highlighting the applicability of our model. 

%% file: discussion.tex
\section{Conclusion}
We have introduced a SPDT graph model and demonstrated its utility for a case study of airborne disease diffusion. The SPDT graph captures both direct and indirect contacts for simulating diffusion process. The proposed graph model is capable of capturing contact dynamics, applying reinforcement for capturing repetitive interactions and public accessibility-based attractiveness to engage in interactions. We demonstrated how the model can be fitted to empirical geo-tag data from social networking App and its ability to reproduce both graph structural properties and diffusion dynamics of empirical graphs. The model generates co-located interaction parameters with considerably low RSE error and reduces prediction error of cumulative spread by up to 95\% over existing graph models. The graph model shows similar response of real graphs to environmental conditions for changing diffusion dynamics. 

The significance of our SPDT model lies in its capture of indirect interactions for diffusion phenomena, which accounts for previously disregarded pathways for transmission. We expect the model to be useful for forecasting infectious disease diffusion within a population given contacts data of a population. More importantly, the model can be used to simulate what-if scenarios to aid health managers and authorities in planning for possible outbreaks and allocating resources for targeted responses. The SPDT graph model can be applied to study diffusion phenomena in on-line social network (OSN) such as on-line post can be seen by current active users instantly while inactive users sees them later on~\cite{gao2015modeling}. This model can also be used to model online and offline activities of users in OSN~\cite{probst2013will}. There are several interesting directions for future developments of our model. Nodes in the current model activate with the same frequency. Thus, it would be interesting to study graph properties with heterogeneous frequencies and find correlation with public accessibility. Another interesting direction is to study the contact set size growth and temporal properties like betweenness. The sensitivity of model parameters to graph stability and model performances against other graph models will also be studied.